\documentclass[12pt]{article}
\usepackage[utf8]{inputenc}
\usepackage[margin=1in]{geometry}

\usepackage{graphicx}
\usepackage{natbib}

\usepackage{amsmath, amsthm, amsfonts, amssymb, latexsym, mathtools}
\usepackage[shortlabels]{enumitem}
\usepackage{bm}
\usepackage{fancyvrb}
\usepackage{longtable}
\usepackage{eurosym}
\usepackage{subcaption}
\usepackage{color}
\usepackage{url}
\usepackage{graphicx}
\usepackage{lscape}
\usepackage{bm}
\usepackage{mathrsfs}

\usepackage{booktabs} 
\usepackage{caption}  

\usepackage{setspace}
\theoremstyle{plain}

\newtheorem{definition}{Definition}

\newtheorem{example}{Example}

\newtheorem{assumption}{Assumption}
\newtheorem{theorem}{Theorem}
\newtheorem{lemma}{Lemma}

\newcommand{\blind}{0}

\begin{document}

\if0\blind
{
  \title{\bf Multivariate quantile regression 
  }

  \author{Antonio F. Galvao\footnote{Department of Economics, Michigan State University. Email: \texttt{agalvao@msu.edu}}
    \and Gabriel Montes-Rojas\footnote{Instituto Interdisciplinario de Economía Política, Universidad de Buenos Aires and CONICET.  Email: \texttt{gabriel.montes@economicas.uba.ar}}
    }
  \maketitle
} \fi

\begin{abstract}
\begin{singlespace}
\noindent This paper introduces a new framework for multivariate quantile regression based on the multivariate distribution function, termed multivariate quantile regression (MQR). In contrast to existing approaches—such as directional quantiles, vector quantile regression, or copula-based methods—MQR defines quantiles through the conditional probability structure of the joint conditional distribution function. The method constructs multivariate quantile curves using sequential univariate quantile regressions derived from conditioning mechanisms, allowing for an intuitive interpretation and flexible estimation of marginal effects. The paper develops  theoretical foundations of MQR, including asymptotic properties of the estimators. Through simulation exercises, the estimator demonstrates robust finite sample performance across different dependence structures. As an empirical application, the MQR framework is applied to the analysis of exchange rate pass-through in Argentina from 2004 to 2024. 
\end{singlespace}

\vspace{4mm}


\vspace{0.25cm}


\end{abstract}

\doublespacing

\newpage

\section{Introduction}

Multivariate distributions lack a natural or intrinsic ordering, making it considerably difficult to extend the concept of quantiles and quantile regression (QR) to multivariate frameworks. Numerous methodologies have been introduced to adapt and broaden the definitions of quantiles and QR for use in multivariate contexts.

A prominent approach utilizes \textit{directional quantiles}, where fixing a direction reduces the multivariate quantile problem to a univariate one. \cite{HPS10} analyze the distributional and quantile characteristics of multivariate response variables using the notion of directional quantiles, building on earlier work such as \cite{Chaudhuri96, Koltchinskii97, Chakraborty03, Wei08}. In this framework, distributional features are explored by examining different directional models. This framework has been applied in various contexts (see, e.g., \cite{PaindeveineSiman11, PaindeveineSiman12, FraimanPateiro12}).
\cite{Montes17, Montes19, Montes22} propose a directional quantile model where the orthonormal basis is fixed—that is, a set of orthogonal directions spans the domain of the dependent variable. In this formulation, directional quantiles in reduced form are defined as the fixed point of a system of directional quantile equations.
A related method is the vector QR. \cite{ChernozhukovGalichonHallinHenry15} and \cite{CarlierChernozhukovGalichon16, CarlierChernozhukovGalichon17} develop a linear vector QR framework that defines a quantile map as the gradient of a convex function, yielding a monotone mapping. 


An alternative approach for multivariate quantile is based on copulas. Since any multivariate distribution can be decomposed into its marginals and a dependence function (copula), copula-based quantile models have been proposed --- see, for example, \cite{BernardaCzado15}. However, such models require strong distributional assumptions about the copula structure.

This paper distinguishes itself from the prior literature by revisiting the fundamental definition of a multivariate quantile, grounded in the distribution function within multivariate settings. The distribution function of a multivariate random variable inherently provides an order to evaluate conditional models. In particular, this corresponds to the potential realization of the random variables being less than (or equal) the given quantile. In other words, quantiles correspond to graphs that represent the same probability of lower value realizations of the random variables involved, offering a valuable measure to assess tail risk, heterogeneity, and scenario valuation. 

Based on the multivariate distribution function, we define a new multivariate quantile regression (MQR) to obtain the marginal effects of covariates. The procedure uses the definition of conditional probability to evaluate a series of conditional univariate QR models, from which the coefficients representing the effects of covariates are obtained, and construct conditional multivariate quantile curves. 

Intuitively, the joint probability of any set of random variables can be expressed as the product of the marginal probability of one variable and the corresponding conditional probabilities of the others using a sequence of conditioning structures. This relationship allows us to break down complex joint distributions into simpler components, making it easier to analyze and understand their behavior. We make use of the conditional expression by evaluating separate conditional QR models.

To build intuition, consider the simple case of a bivariate model, say $Y_1$ and $Y_2$, for which we want to evaluate the $\tau$-quantile MQR model conditional on a set of covariates $X$. Using the decomposition of the joint probability into conditional and marginal probabilities, we break down a single quantile $\tau=P[Y_1\leq q_1^\tau(X),Y_2\leq q_2^\tau(X)|X]$ into $\tau=\tau_{2|1}\times \tau_1$, with $\tau_1=P[Y_1\leq q_1^\tau(X)|X]$ and $\tau_{2|1}=P[Y_2\leq q_2^\tau(X)|X,Y_1\leq q_1^\tau(X)]$. Then, define  the $\tau$-MQR given by $(q_1^\tau(X),q_2^\tau(X))$ as the associated quantiles of: (i) the univariate $\tau_1$ QR for $Y_1$ on $X$; (ii) the $\tau_{2|1}$ univariate QR coefficients of $Y_2$ on $X$, also conditional on $Y_1\leq q_1^{\tau}(X)$. Note that, in the multivariate setting, $(q_1^\tau(X),q_2^\tau(X))$ is a graph or curve rather than a point estimate. Thus, we are particularly interested in the marginal effect that a change in an element $k$ of $X$ produces on this graph or curve, e.g. $(\partial q_1^\tau(x)/\partial x_{k},\partial q_2^\tau(x)/\partial x_{k})$, if $X_k$ is continuous, or e.g. $(q_1^\tau(x_k=1,x_{-k})-q_1^\tau(x_k=0,x_{-k}),q_2^\tau(x_k=1,x_{-k})-q_2^\tau(x_k=0,x_{-k}))$ if $X_k$ is binary.

Given this decomposition, we propose a sequential estimator for the conditional effects of interest based on univariate QR models. The practical implementation is simple and based on standard univariate QR estimation. We then establish the limiting asymptotic properties of the estimator. In particular, the estimated coefficients' properties are studied based on linear QR asymptotic properties using the Bahadur representation with generated regressors.

A natural question is: when are such models of interest? The MQR model is of interest for any case where the empirical researcher is interested in evaluating the (marginal) impact of covariates on the joint distribution of random variables. An important application involves treatment effect analyzes of multivariate continuous outcomes, such as the effect of a drug on multiple health variables. Moreover, this is of interest for economic models that involve trade-offs of outcome variables. 
Consider the following examples from the economics literature.

\vspace{.5cm}
\noindent \textbf{Monetary policy.} Suppose a central bank (e.g., the Federal Reserve) models the joint behavior of three variables: $(r, y, p)$, where $r$ is the policy interest rate, $y$ represents GDP (level, growth, or output gap), and $p$ is inflation. The central bank typically faces a trade-off between stabilizing output and inflation, the so-called Phillips curve relationship linking both variables. In this context, it is important to model "good" or "bad" scenarios—e.g., both inflation and output are unusually high or low by examining joint probability quantiles. The MQR allows to evaluate the effect of a change in $r$ on the joint distribution of $(y,p)$, thus evaluating heterogeneity in monetary policy.

\vspace{.5cm}

\noindent  \textbf{Exchange rate pass-through.} A government may consider a model with $(e, y, p)$, where $e$ is the change in the exchange rate (depreciation/appreciation), $y$ again denotes GDP, and $p$ is inflation. The policymaker’s goal might be to understand how extreme movements in the exchange rate jointly affect output and inflation. Here again, MQR can be used to analyze tail behavior and policy trade-offs.

\vspace{.5cm}

The remainder of the paper is organized as follows. Section \ref{bivariate} introduces the bivariate model and highlights the core intuition behind the proposed approach. Section \ref{multivariate} extends the framework to the general multivariate case. Section \ref{comparison} discusses how the proposed method compares with existing approaches to multivariate quantile modeling. Section \ref{application} presents an empirical application to exchange rate pass-through in Argentina. Finally, Section \ref{conclusion} offers concluding remarks. Mathematical proofs are gathered in the Appendix.

\section{Motivation and the bivariate case}\label{bivariate}

\subsection{The unconditional case}
\begin{figure}
\centering
    \caption{Univariate case}
    \label{fig:1}
    \begin{tabular}{cc}
        \includegraphics[scale=0.5]{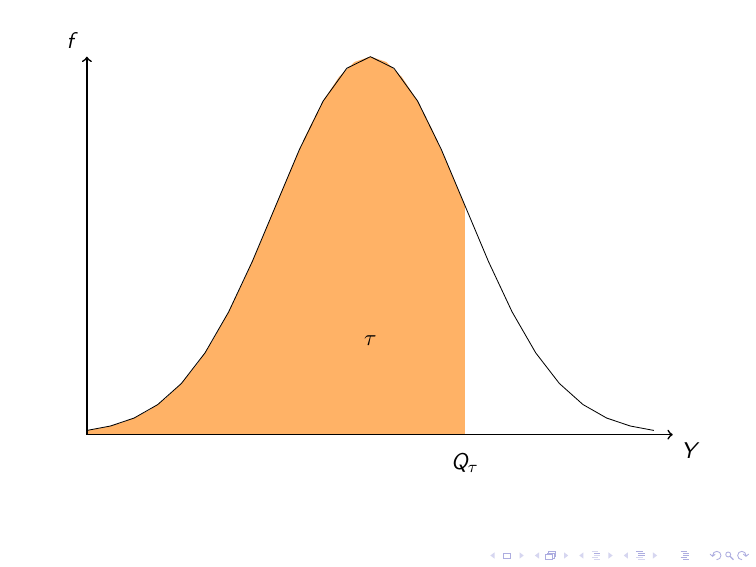}&
        \includegraphics[scale=0.5]{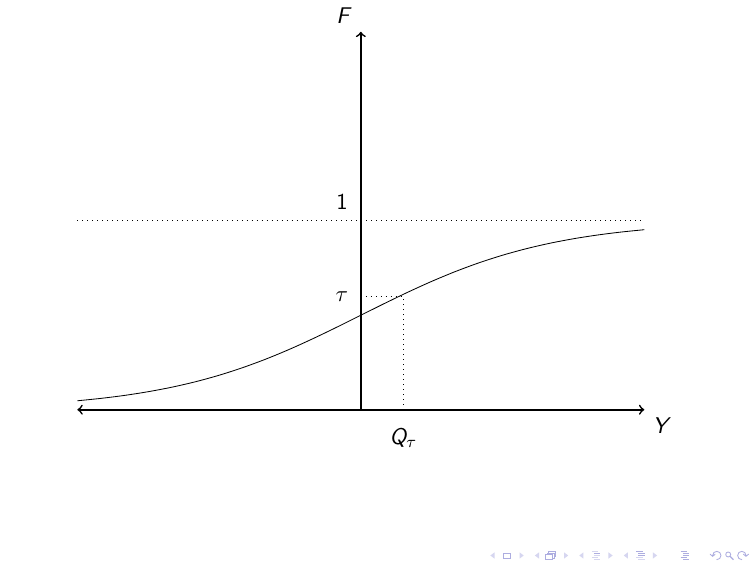}
    \end{tabular}
\end{figure}

Consider a univariate random variable \( Y \) with domain \( \mathcal{Y} \subseteq \mathbb{R} \) and conditional distribution function (CDF) \( F(y) := P(Y \leq y) \). 
Then, the \( \tau \)-th quantile for \( \tau \in (0,1) \) is defined as \( Q_\tau(Y) := \inf \{ y \in \mathcal{Y} : \tau \leq F(y) \} \). Note that if \( F(\cdot) \) is continuous, then \( Q_\tau(Y) = F^{-1}(\tau)\). For many cases, the quantile is assumed to be unique, and as such it is well defined. Figure \ref{fig:1} illustrates the basic statistical concepts.

\begin{figure}[ht]
    \centering
    \caption{(a) Two points representing quantile $\tau$. $P(Y_1 \leq q_1, Y_2  \leq q_2) =\tau$. (b) Two contour lines for $\tau<\tau'$}
    \label{fig:biv} 
    \centering
    \begin{tabular}{cc}
    (a) & (b) \\
        \includegraphics[scale=0.6]{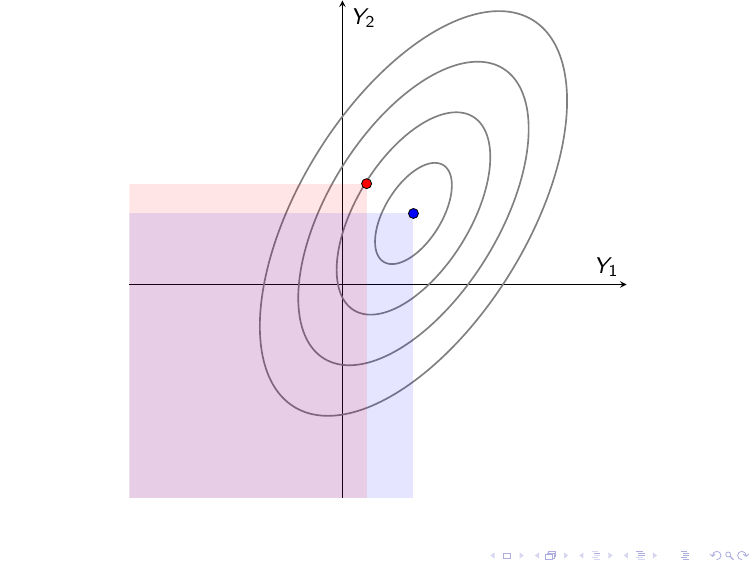}&
        \includegraphics[scale=0.6]{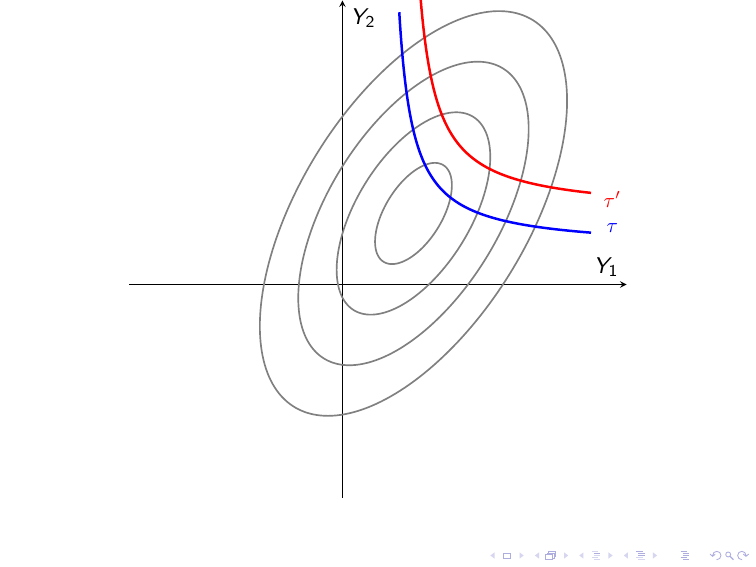}
    \end{tabular}
\end{figure}

However, for a multivariate random variable, say \( Y =(Y_1,Y_2)\) with domain \( \mathcal{Y} \subseteq \mathbb{R}^2 \), \( \inf \{ y \in \mathcal{Y} : \tau \leq F(y) \} \) is (in general) not unique. Quantiles are themselves then defined on regions, contours and depths \citep{HallinKonen24}. Consider the following definition
\begin{definition}
A quantile on the bivariate domain is any pair $(q_1,q_2)$ such that $P(Y_1 \leq q_1, Y_2 \leq q_2)=\tau$. 
\end{definition}
Figure \ref{fig:biv}a plots the contour plot for the probability density function (PDF) of a bivariate distribution and adds two points that correspond to the same quantile $\tau$. Figure \ref{fig:biv}b plots the same contour plot but considers the curves corresponding to two different $\tau$'s.

For motivation, we develop the idea for the bivariate case, its extension for general multivariate cases is presented in the next section. First, we are interested in a model for a given quantile $\tau$ as in Figures \ref{fig:biv}a and \ref{fig:biv}b. Second, the main idea follows from the standard definition of joint probability,
\begin{equation}\label{eq:motivation}
P(Y_1 \leq q_1^\tau, Y_2 \leq q_2^\tau)=P(Y_2 \leq q_2^\tau| Y_1 \leq q_1^\tau)P(Y_1 \leq q_1^\tau).
\end{equation}

This simply means that for any given $\tau$ on the left-hand side of \eqref{eq:motivation}, there are two possible quantile indices (i.e. probabilities) such that
\begin{equation*}
\tau=\tau_{2|1}\tau_1,
\end{equation*}
where $\tau_1=P(Y_1 \leq q_1^\tau)$ represents a quantile on the marginal distribution of $Y_1$ (which is not necessarily $\tau$, but $Q_{\tau_1}(Y_1)$, the $\tau_1$ quantile of $Y_1$), and $\tau_{2|1}=P(Y_2 \leq q_2^\tau| Y_1 \leq q_1^\tau)$ with $q_2^\tau$ is the $\tau_{2|1}$-quantile of the conditional distribution of $Y_2|(Y_1\leq q_1^\tau)$. Here $(q_1^\tau,q_2^\tau)$ is a graph or curve rather than a point estimate, as illustrated in Figure \ref{fig:biv}b.

Equation \eqref{eq:motivation} contains two important features. First, it decomposes the joint CDF, which defines the multivariate quantile, into two univariate CDFs, which are associated with simple univariate quantiles. Second, for a given quantile $\tau$ of interest for the multivariate case, we can find the quantities $q_1^\tau$ and $q_2^\tau$ where each is associated with a univariate quantile, that is $\tau_{2|1}$ and $\tau_1$, respectively.
Note that we require $\tau_1\geq \tau$ and $\tau_{2|1}=\tau/\tau_2$.
The reverse is, of course, also true where we condition on $Y_2$ first and then we consider the model for $Y_1$ conditioning on $Y_2$. For this case we would use $\tau=\tau_{1|2}\tau_2$. This implies that the univariate quantiles could be implemented by a particular order or permutation, $1\rightarrow 2$ and $2\rightarrow 1$. The next example illustrates the decomposition of a bivariate case into univariate, as well as the corresponding choice of quantile parameters.

\begin{example}
Consider a simple example of a bivariate normal distribution,

\[
\begin{pmatrix}
Y_1 \\
Y_2
\end{pmatrix} \sim \mathcal{N} \left( \begin{pmatrix}
\mu_1 \\
\mu_2
\end{pmatrix}, \begin{pmatrix}
\sigma_1^2 & \rho \sigma_1 \sigma_2 \\
\rho \sigma_1 \sigma_2 & \sigma_2^2
\end{pmatrix} \right),
\]
where:
$\mu_1, \mu_2$ are the means of $Y_1$ and $Y_2$, $\sigma_1^2$ and $\sigma_2^2$ are the variances of $Y_1$ and $Y_2$,
$\rho$ is the correlation between $Y_1$ and $Y_2$. 
The conditional distribution of \(Y_2\) given \(Y_1 = y_1\) follows a normal distribution:
\begin{equation*}
Y_{2}\mid Y_{1}=y_{1}\ \sim \ \mathcal{N}\left(\mu_{2}+{\frac{\sigma_{2}}{\sigma_{1}}}\rho (y_{1}-\mu_{1}),\,(1-\rho^{2})\sigma_{2}^{2}\right).
\end{equation*}
Then, we fix quantile $\tau_{1}$ and calculate this quantile, $q_1^{\tau_1}$, of $Y_{1}$ as $q_1^{\tau_1} =\mu_{1} +\sigma_{1} \sqrt {2}\Phi^{-1}(2\tau_{1}-1)$ where $\Phi()$ is the univariate standard normal cdf. 
Next, we are able to write the conditional distribution $Y_{2}|Y_{1}=q_1^{\tau_1}$ as 
\begin{align*}
Y_{2}\mid Y_{1} = q_1^{\tau_1} & \sim \ \mathcal{N}\left(\mu_{2}+{\frac{\sigma_{2}}{\sigma_{1}}}\rho (q_1^{\tau_1}-\mu_{1}),\,(1-\rho^{2})\sigma_{2}^{2}\right)\\
& = \mathcal{N}\left(\mu_{2}+{\frac{\sigma_{2}}{\sigma_{1}}}\rho \left(\mu_{1} +\sigma_{1} {\sqrt {2}}\Phi^{-1}(2\tau_{1}-1)-\mu_{1} \right),\,(1-\rho^{2})\sigma_{2}^{2}\right)\\
& = \mathcal{N}\left(\mu_{2}+\sigma_{2}\rho  \sqrt {2}\Phi^{-1}(2\tau_{1}-1) ,\,(1-\rho^{2})\sigma_{2}^{2}\right).
\end{align*}
Finally, we are able to compute the $\tau_{2}$-quantile of $Y_{2}|Y_{1}=q_1^{\tau_1}$ as
\begin{equation*}
\left[ \mu_{2}+\sigma_{2}\rho \sqrt {2}\Phi^{-1}(2\tau_{1}-1) \right] + \left( (1-\rho^{2})\sigma_{2}^{2} \right) \sqrt {2}\Phi^{-1}(2\tau_{2}-1).
\end{equation*}

We now seek the distribution of \(Y_2\) given that \(Y_1 \leq q_1^{\tau_1}\). This conditional distribution is still Gaussian but truncated, from which the quantile $\tau_{2|1}$ can be obtained, say $q_2^{\tau_{2|1}}$. The parameters of this distribution are as follows. (i) The conditional mean is $
\mu_{Y_2 | Y_1 \leq q_1^{\tau_1}} = \mu_2 + \frac{\rho \sigma_2}{\sigma_1} E[Y_1 | Y_1 \leq q_1^{\tau_1}] - \mu_1$, and (ii) the conditional variance is $
\sigma^2_{Y_2 | Y_1 \leq q_1^{\tau_1}} = \sigma_2^2 (1 - \rho^2) \left( 1 - \frac{ \phi\left( \frac{q_1^{\tau_1} - \mu_1}{\sigma_1} \right) }{ \Phi\left( \frac{q_1^{\tau_1} - \mu_1}{\sigma_1} \right) } \right)$,
where \(\phi\) is the standard normal pdf.

Finally, for any $\tau\in(0,1)$, if we take $\tau_1\geq \tau$ and $\tau_{2|1}=\tau/\tau_1$, we can use the truncated distribution above to solve for the $\tau$-multivariate quantile graph $(q_1^{\tau}=q_1^{\tau_1},q_2^{\tau}=q_2^{\tau_{2|1}})$.
\end{example}

\subsection{The regression case}
Now we extend the previous ideas to a conditional model in which we have covariates. The main object of interest in this paper is to compute the partial effect of a regressor on the multivariate conditional quantile function.

\begin{figure}
\begin{center}
    \caption{Effects of a change in $X$ from $x$ to $x'$}
    \label{fig:4}
    \includegraphics[width=8cm]{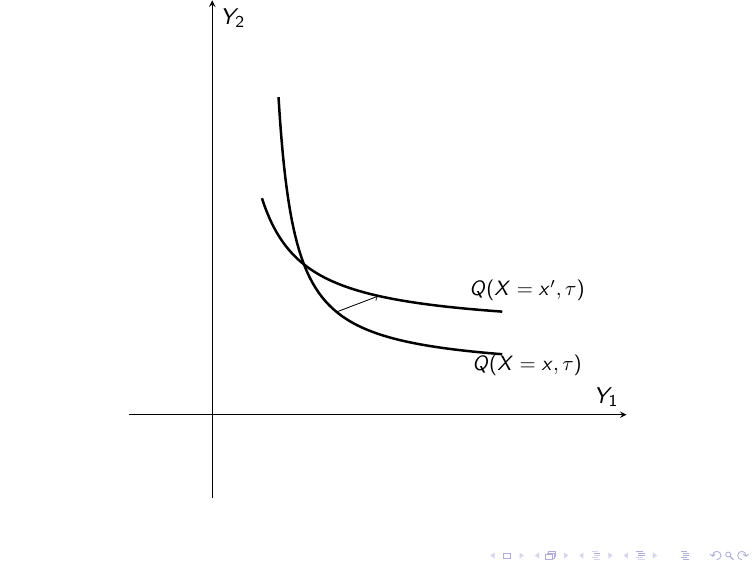}
\end{center}
\end{figure}

Consider a $k$-dimensional vector of regressors $X$ with domain in $\mathcal{X} \subseteq \mathbb{R}^k$, which represents the explanatory or control variables. 
The main goal is to obtain the effects of a marginal change in covariates, as for example in Figure \ref{fig:4}. Note that distribution function-quantile curves can cross each other. That is, the covariates may have many different effects on the bivariate distribution function, and thus on the corresponding estimated quantile curves.

Consider the following extension of \eqref{eq:motivation} to the conditional probability case 
\begin{equation*}
    P(Y_1 \leq q_1^\tau, Y_2 \leq q_2^\tau|X)=P(Y_2 \leq q_2^\tau| Y_1 \leq q_1^\tau,X)P(Y_1 \leq q_1^\tau|X).
\end{equation*}
The bivariate quantile map of interest is 
\begin{equation*}
Q^{1\rightarrow 2}:=(q_1,q_2) : \mathcal{X} \times [0,1] \mapsto \mathcal{Y},
\end{equation*}
and corresponds to our proposed definition of multivariate quantile regression (MQR).

Thus, to obtain the $\tau$-quantile bivariate QR model using the permutation $1\rightarrow 2$, we consider the following procedure for a grid of $\tau_1\in(\tau,1)$:
\begin{enumerate}
    \item Estimate the univariate QR model for $q_1^\tau(X):=Q_{\tau_1}(Y_1|X)=\beta_1(\tau_1)X$ for a given $\tau_1\in(\tau,1)$. This is a standard conditional QR model.
   \item Estimate the univariate quantile regression model for $q_{2|1}^\tau(X):=Q_{\tau_{2|1}}(Y_2|Y_1\leq  q_1^\tau(X),X)=\beta_{2|1}(\tau_{2|1})X$ for  $\tau_{2|1}=\tau/\tau_1$. This is a standard conditional QR model for all observations that satisfy $Y_1\leq  q_1^\tau(X)$.
    \item[2'] An alternative numerically equivalent step 2 is to use the ``generated regressor" $D^{\tau_1}:=I(Y_1\leq  q_1^\tau(X))$. That is, $q_{2|1}^\tau(X)=\beta_{2|1}(\tau_{2|1})D^{\tau_1}X$ are the first components in $Q_{\tau_{2|1}}(Y_2|Y_1\leq  q_1^\tau(X),X)=\beta_{2|1}(\tau_{2|1})D^{\tau_1}X+\beta^-_{2|1}(\tau_{2|1})(1-D^{\tau_1})X$.
\end{enumerate}
Step 2 and Step 2' above deliver the exact same parameter estimator for $\beta_{2|1}(\tau_{2|1})$. The latter has the advantage that the asymptotic theory can be developed using two-step M-estimator models for generated regressors.

This delivers $B^{1\rightarrow 2}(\tau)=[\beta_{1}(\tau)^\top:=\beta_{1}(\tau_{1})^\top,\beta_{2}(\tau)^\top:=\beta_{2|1}(\tau_{2|1})^\top]^\top$, where $\top$ denotes the transpose. Then, the quantile graph is given by $B^{^{1\rightarrow 2}}(\tau)X$ of the form
\begin{equation*}
   Q^{1\rightarrow 2}(X,\tau)=(q^{\tau}_{1}(X),q^{\tau}_{2|1}(X))^\top.
\end{equation*}

A similar exercise can be developed for the other case, $2\rightarrow 1$, using $\tau_2\in(\tau,1)$ to obtain
$B^{2\rightarrow 1}(\tau)=[\beta_{1}(\tau)^\top:=\beta_{1|2}(\tau_{1|2})^\top,\beta_{2}(\tau)^\top:=\beta_{2}(\tau_{2})^\top]^\top$ and $Q^{2\rightarrow 1}(X,\tau)$. It should be noted that both permutation procedures deliver alternative estimates of the same bivariate model. The choice of which permutation to use can be based on a selection of the most appropriate specification for the marginal models, which in the end is similar to a Cholesky decomposition in time-series models. However, if the model is correctly specified in each case, both permutations should deliver similar estimates. This can be checked by comparing estimates across permutations.\footnote{We investigate the effect of this choice in finite samples in Section \ref{sec:MC} below. Results show evidence that when the model is well specified, the choice does not affect results.}

To illustrate, consider the following simulation exercises for a bivariate distribution $(Y_1,Y_2)$:
\begin{align*}
    Y_1&=\beta_{10}+\beta_{11} X + (1+\alpha_{1}|X|)\varepsilon_{1},\\
    Y_2&=\beta_{20}+\beta_{21} X +     \gamma_{21} Y_1+ (1+\alpha_{2}|X|)\varepsilon_{2}.   
\end{align*}
Here we assume that $X$, $\varepsilon_{1}$ and $\varepsilon_{1}$ are independent $\mathcal{N}(0,1)$, $\beta_{10}=\beta_{11}=\beta_{20}=\beta_{21}=\gamma_{21}=1$. Then, conditional on $X$ we obtain a bivariate normal distribution of $(Y_1,Y_2)$.

\begin{figure}[ht]
    \centering
      \caption{Comparison of estimation and simulation for bivariate models: (a) location shift only, and (b) location-scale shift.}
    \label{fig:comparison} 
    \begin{subfigure}[b]{0.49\textwidth}  
        \centering
        \includegraphics[width=\textwidth]{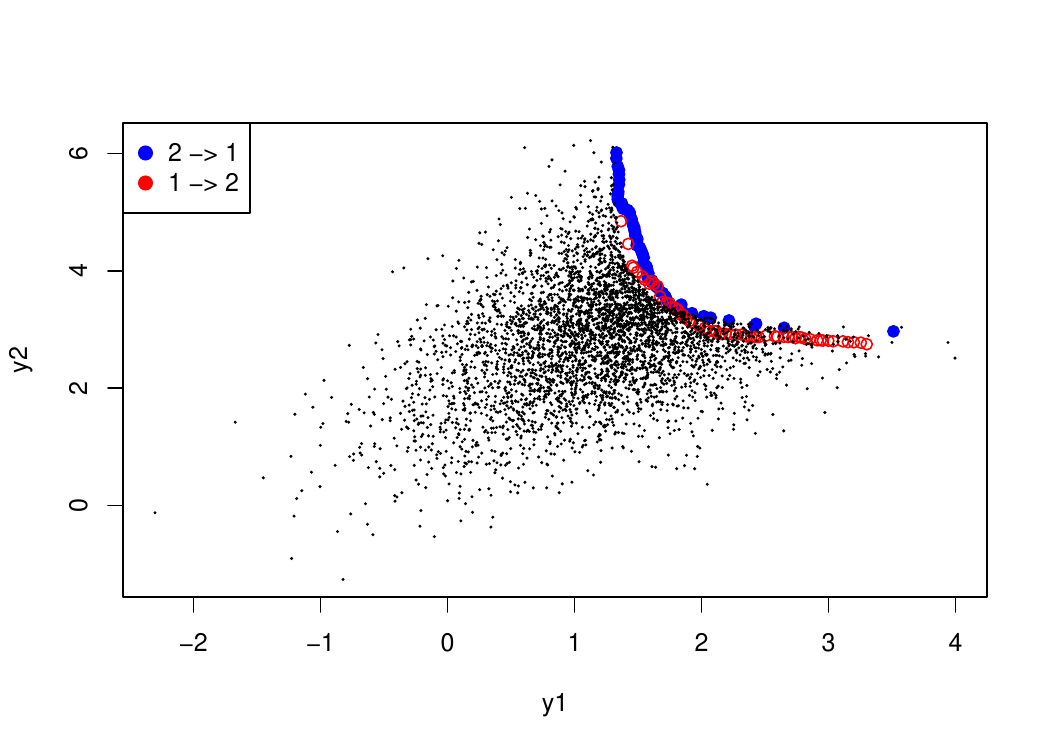}
        \caption{Estimation and simulation for location shift only bivariate model. $X=1$}
        \label{fig:a0}
    \end{subfigure}
    \hspace{0.02\textwidth}  
    \begin{subfigure}[b]{0.49\textwidth}  
        \centering
        \includegraphics[width=\textwidth]{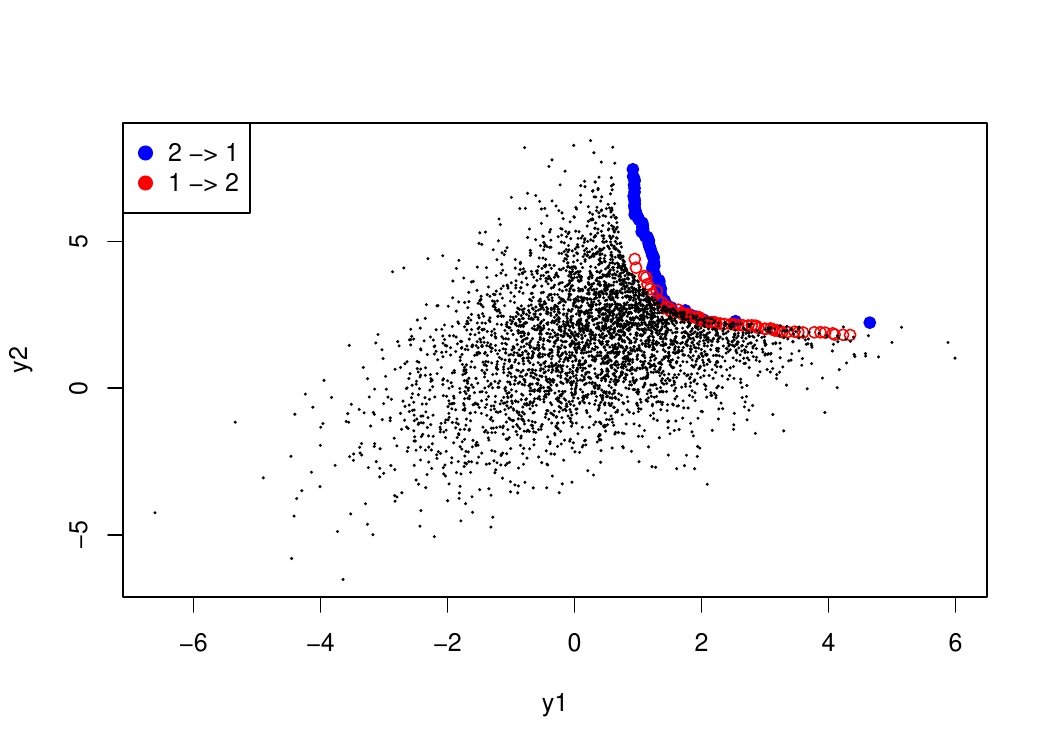}
        \caption{Estimation and simulation for location-scale shift bivariate model. $X=1$}
        \label{fig:a1}
    \end{subfigure}
\end{figure}

We consider two different simulations for $n=1000$ in Figures \ref{fig:a0} and \ref{fig:a1} for $\tau=0.25$. We use a quantile grid of 0.01 spacing and plot the estimated quantile curves for both conditioning models. In each figure the blue dots correspond to the estimation with $\tau_2$ first ($2\rightarrow 1$ permutation), and the red ones to those starting with $\tau_1$ ($1\rightarrow 2$ permutation). 

The first one uses $\alpha_{1}=\alpha_{2}=0$ and it corresponds to location shifts of the covariate $X$ in each variable. The second one uses $\alpha_{1}=\alpha_{2}=1$ and it is a location-scale shift model in both variables. Figures \ref{fig:a0} and\ref{fig:a1} present the quantile curves estimation for $X=1$ for the first and second cases above, together with a random draw of 10,000 of the corresponding bivariate normal distribution for a fixed value of covariate, with the points that satisfy the quantile condition. 

Note that both models deliver a good approximation to the effect of bivariate quantile graph of level $\tau=0.25$ for $X=1$. However, the specification $1\rightarrow 2$ works better as the linear specification is more appropriate.

\section{The general multivariate model}\label{multivariate}

\subsection{The model}

Consider an $m$-dimensional process \( Y = (Y_{1}, ..., Y_{m})^\top\), with domain in \( \mathcal{Y} \subseteq \mathbb{R}^m \), representing the endogenous variables. Additionally, consider a $k$-dimensional process \( X \) with domain in \( \mathcal{X} \subseteq \mathbb{R}^k \), which represents the explanatory or control variables. 

Of particular interest is the case where the covariates are generated by the \( \sigma \)-field generated by \( \{Y_s, s \leq t\} \) and all other available information at time \( t \), denoted by \( \mathcal{F}_t \). In this scenario, the model becomes a quantile autoregressive regression model (VARQ) as in \cite{Montes19}. For an autoregressive model of order \( p \), the process \( X_{t-1} \) is defined as $X_{t} = [1,Y_{t-1}^\top, Y_{t-2}^\top, \dots, Y_{t-p}^\top]^\top$, where \( k = m \times p \).

The main goal is to obtain a conditional model for $Y|X$ that delivers a semiparametric model for the conditional multivariate quantile regression (MQR).

\begin{definition}
    Quantiles are analyzed in terms of the multivariate linear regression
\begin{equation}
    \tau=P\left[Y_{1}\leq X \beta_1(\tau), Y_{2}\leq X \beta_2(\tau),...,Y_{m}\leq X \beta_m(\tau)|X\right],\ \tau\in(0,1),
\end{equation}
where \( \beta_j(\tau) \), for \( j = 1, 2, \dots, m \), are the corresponding \( k \times 1 \) coefficient vectors for the \( j \)-th element of \( Y \). Define
\begin{equation}
Q_{Y | X}(\tau | X)=[q_1^\tau(X),...,q_m^\tau(X)]^\top = B^\top(\tau)X^\top ,\ \tau\in(0,1),
\end{equation}
where $q_j^\tau(X)=X\beta_j(\tau)$ and $B(\tau) \equiv [\beta_1(\tau), \beta_2(\tau), \dots, \beta_m(\tau)]$
is a vector of coefficients of size \( k \times m \).
 The quantile graph \( Q (X,\tau)\) is defined as $Q: \mathcal{X} \times (0,1) \mapsto \mathcal{Y}$,
and corresponds to our proposed definition of multivariate quantile regression (MQR).
\end{definition}

The MQR estimates require an order or permutation among the endogenous variables for implementation. In particular, there are $m!$ possibilities to map the MQR models -- for the binary case there are only $2!=2$ possibilities, $1\rightarrow 2$ and $2\rightarrow1$. Let $(p)$ denote a given permutation. For notation purposes, we do not index all parameters by $(p)$ in what follows.

Given a permutation $(p)$, the sequential conditioning probability formula is
\begin{equation}
\tau=\Pi_{j=2}^{m}P\left(\cap_{l=j}^{m} (Y_l\leq q_l^\tau(X))|\cap_{h=1}^{l-1} (Y_h\leq q_h^\tau(X)),X\right)P(Y_1\leq q_1^\tau(X)|X),
\end{equation}
where $q_j^\tau(X):=X \beta_j(\tau)$.
Then, we need to consider a sequence of univariate QR models $\{\tau_1,\tau_{2|1},\tau_{3|(1,2)},...,\tau_{m|(1,2,...,m-1)}\}$ that satisfy the sequence of restrictions $\tau_1\in(\tau,1)$, $\tau_{2|1}\in(\tau/\tau_1,1)$, $\tau_{3|(1,2)}\in(\tau/(\tau_1\tau_{2|1}),1)$, ..., $\tau_{m|(1,...,m-1)}\in(\tau/(\tau_1\tau_{2|1}...\tau_{m-1|1,2,...,m-2)}),1)$. For simplicity, in what follows, we label this vector as $(\tau_1,\tau_{2},...,\tau_{m-1},\tau_{m})$, and the permutation order is omitted. Moreover, this sequence must satisfy $\tau=\Pi_{j=1}^m\tau_j$ in all cases.

\subsection{A sequential MQR estimator}

For a given permutation $(p)$ and for each $j=2,...,m$, define $d_{ji}=I[y_{ji}\leq q_j^{\tau_j}(x_i)]$. Moreover also for $j=2,...,m$,
define 
\begin{equation*}
w_{ji}:=\left\{(d_{1i},d_{2i},...,d_{j-1i}) \mid d_{1i},d_{2i},...,d_{j-1i}\in\{0,1\}\right\},
\end{equation*}
that is, all the possible conditioning restrictions. Now define 
$z_{ji}:=(x_{i}\otimes w_{ji})$. For variable $1$ let $z_{1i}:=x_{i}$. Note that the dimensions of $z_j$ are $k_j:=k\times 2^{j-1}$.

The variables $d$ used to construct $w$ in $z$ are themselves generated regressors. Define $\hat z_{ji}$ as the regressors generated by interacting all the possible combinations of  $\hat d_{hi}:=I(y_{hi}\leq \hat z_{hi} \hat\beta_h^+(\tau_h))$ and $(1-\hat d_{hi})$ for $h=1,2,...,j-1$ with $x_{i}$, and the $\hat\beta_h^+$ estimators being defined next. 

Then, the estimator of the $j$-th equation of the MQR model is given by the  QR model for $\tau_j$ defined by
\begin{equation}\label{eq:qr}
\hat{\beta}_j^+(\tau_j) := \arg \underset{\beta \in \mathbb{R}^{k_j}}{\min} \frac{1}{n} \sum_{t=1}^n \rho_{\tau}(y_{ji} - \hat z_{ji}\beta),
\end{equation}
where $\rho_{\tau}(u) := u \{  \tau - I( u \leq 0) \}$. Define $B^{+}(\tau)=(\beta_1^+(\tau_1)^{\top},...,\beta_m(\tau_m)^{\top})^\top$ as the collection of all QR coefficients and $\hat B^{+}(\tau)$ its estimator.

The first $k$ elements in the $k_j \times 1$ vector $\hat{\beta}_j^+(\tau_j)$ are the MQR coefficients  for the $j$-th component (i.e., the ones that correspond to the conditioning restrictions). Call this estimator $\hat{\beta}_j(\tau_j)$.
Note that we do not have the same covariates for all $j=1,2,...,m$ endogenous variables, and this varies depending on the permutation used to estimate the MQR. Moreover, this is a sequential estimator, where for each $j$ equation the previous $m-j$ models where estimated (for the $m$th equation no previous model was used).

\begin{definition}
The $\tau$-MQR coefficients are given by the first $k$ elements of the QR coefficients obtained from \eqref{eq:qr}, $\hat{B}(\tau):=[\hat{\beta}_1(\tau_1),\ldots,\hat{\beta}_m(\tau_m)]$, where $\tau=\Pi_{j=1}^m\tau_j$. Moreover, the estimated quantile graph is given by $\hat{Q}(X,\tau)=\hat{B}^\top(\tau)X^\top$.  
\end{definition}

\subsection{Asymptotic properties of the MQR estimator}
\label{sec: null}

In this subsection, we derive the limiting distribution of the sequential MQR estimator. To this end, we consider first the limiting distribution of the estimators 
$\hat{\beta}_j(\tau_j)$ over $\mathcal{T}\in(0,1)$ for all, $j=1,2,...,m$, as defined previously, assuming that the conditioning process that generates the covariates $z_{ji}$ is known and based on the true quantile function. Then, we evaluate the impact of the generated regressors $\hat{z}_{ji}$.

Here we impose the following regularity conditions.

\begin{assumption} \label{assump1}The following conditions apply:

    \begin{itemize}
       	\item[(A1)] $\{ (y_{1i},y_{2i},...,y_{mi}.x_{i}), \ i=1,2,...\}_{i=1}^n$  is a random sample of independent  vectors that satisfies $E[\| ({y}_{i},x_{i}) \|^{\phi}] < \infty$ for some $\phi > 2$.
	\item[(A2)]
	For all $j=1,2,...,m$, let $F_j(\cdot | {z})$ denote the conditional distribution function of $u_{ji}=y_{ji}-z_{ji}\beta_j^+(\tau_j)$ given $z_{ji}=z$, for all $j=1,2,...,m$.  Assume that  $F_j(\cdot | {z})$ has a Lebesgue density $f_j(\cdot | {z})$ such that
	\begin{enumerate}
		\item[(i)]  $| f_j(. | {z}) | \leq C_{f}$ on the support of $(y_{j},z_{j})$ for some constant $0<C_{f}<\infty$.
		\item[(ii)] $| f_j(y_1 | z) - f_j(y_2 | z) | \to 0$ as $| y_1 - y_2 | \to 0$ for each fixed $z$.
        \item[(iii)] Assume that $F_{1,...,m}$, the joint distribution function of $(Y_1,...,Y_m)|X$ conditional on $X$, is absolutely continuously differentiable on all its arguments in the entire domain $\mathcal{Y}$. All its derivatives are dominated by the same finite constant $C_f$ as in (i).
	\end{enumerate}
	\item[(A3)] For all $j=1,2,...,m$, there exist an open set $\mathcal{T}^{*} \subset (0,1)$ with $\mathcal{T}^{*} \supset \mathcal{T}$ such that for each $\tau \in \mathcal{T}^{*}$, and for all $j=1,2,...,m$, there exists a unique vector ${\beta}_j^+(\tau) \in \mathbb{R}^{k_j}$ that solves the equation
	$E [ ( \tau - I\{ y_{ji} \leq z_{ji}\beta_{j}^+(\tau) \} )z_{ji}^\top ] = 0$.
	\item[(A4)] For all $j=1,2,...,m$, define the matrices
	\begin{equation*}
	\Omega_{0j}:= \mbox{plim}_{n\rightarrow\infty}  \frac{1}{n}\sum_{i=1}^nz_{ji}^\top z_{ji},\ j=1,2,...,m,
    \end{equation*}
    \begin{equation*}
	\Omega_{1j}(\tau_j) := \mbox{plim}_{n\rightarrow\infty}  \frac{1}{n}\sum_{i=1}^n f_j(0 | z_{ji}) z_{ji}^\top z_{ji}, \ j=1,2,...,m.
	\end{equation*}
	Assume that $\Omega_{0j}$ is positive definite, and 	$\Omega_{j1}(\tau_j)$ is positive definite for each $\tau \in \mathcal{T} $.
\end{itemize}
\end{assumption}

These conditions are standard in the QR literature. Condition A1 is a random sample assumption. This includes $iid$ as special case. The analysis that follows also works for strict stationary and $\beta$-mixing as in \cite{GalvaoKatoMontesOlmo14}, although the derivation of the variance-covariance matrix of the estimator requires further development. 
Condition A2 and A3  are standard on the conditional densities and conditional moments, respectively. Condition A3-iii is special for the multivariate analysis.
Condition A4 guarantees that the matrices $\Omega_{0j}$ and $\Omega_{1j}(\tau_j)$ do not degenerate for each  $\tau \in \mathcal{T}$. 

As discussed in \cite{GalvaoKatoMontesOlmo14}, and because of the computational property of the QR estimate -- Theorem 3.1 in \cite{KoenkerBassett78} --, we can select $\hat{\beta}_j^+(\tau_j)$ in such a way that
the path $\tau_j \mapsto \hat{\beta}_j^+(\tau_j)$ is bounded and continuous. Therefore, we may assume that the path $\tau_j \mapsto \sqrt{n} \{ \hat{\beta}_j^+(\tau_j) - \beta_j^+(\tau_j) \}$ is bounded
over $\tau_j \in \mathcal{T}$.
Let $\ell^{\infty}(\mathcal{T})$ denote the space of all bounded functions on $\mathcal{T} $ equipped with
the uniform topology, and let $(\ell^{\infty}(\mathcal{T}))^{k_j}$ denote the $k_j$-product space of $\ell^{\infty}(\mathcal{T})$ equipped with the product topology. We use the notation $\Rightarrow$ for weak convergence. For $a,b \in \mathbb{R}$, we write $a \wedge b = \min \{ a,b \}$.

\begin{lemma}
	\label{thm1}
	Assume that conditions (A1)-(A4) in Assumption 1 hold.
	Then, for each $j=1,2,...,m$, $\beta_j^+(\tau_j)$ admits the following Bahadur representation
	\begin{equation}
	\sqrt{n} \{ \hat{\beta}_j^+(\tau_j) - \beta_j^+(\tau_j) \} =  \Omega_{1j}(\tau_j)^{-1} \frac{1}{\sqrt{n}} \sum_{i=1}^n  \left \{ \tau_j - I  [y_{ji} \leq z_{ji}\beta_{j}^+(\tau_j) ]  \right \} z_{ji}^\top
	+ o_p(1), \label{eq:bahadur}
	\end{equation}
	for $\tau_j\in\mathcal{T}$, $n\rightarrow\infty$. Therefore, it follows that
	\begin{equation}
	\sqrt{n} \{ \hat{\beta}_j^+(\tau_j) - \beta_j^+(\tau_j) \}
	\Rightarrow \Gamma_j^+(\tau_j):=\Omega_{j1}(\tau_j)^{-1} W_j(\tau_j) \ \text{in} \ (\ell^{\infty}(\mathcal{T}))^{k_j}, \label{eq:asymp}
	\end{equation}
	where $W_j(\tau_j)$ is a zero-mean, continuous Gaussian process on $\mathcal{T}$ with covariance kernel
	\begin{equation}
	E [ W_j(\tau_{j}) W_j(\tau'_{j})^\top ] =  (\tau_{j}
	\wedge \tau'_{j}-\tau_{j} \tau'_{j}) \Omega_{0j}.
	\label{eq:asympker}
	\end{equation}
\end{lemma}

Lemma \ref{thm1} shows weak convergence uniformly over $\mathcal{T}$ of the estimator of a single equation for the MQR model. The joint limiting distribution of the full $m$-variate model of the $(p)$ permutation can then be obtained by noting that $\hat{B}(\tau)$ is constructed as the collection of the $m$ QR coefficients corresponding to the sequential conditioning mechanism, each containing the first $k$ elements. Then, from Lemma \ref{thm1}
\begin{equation}\label{eq:asympB}
    \sqrt{n} \{ \hat{B}(\tau) - B(\tau) \}
	\Rightarrow \Gamma(\tau),
\end{equation}
where $\Gamma(\tau)$ corresponds to a subset of the joint limiting distribution of the processes, say $\{\Gamma_1^{+}(\tau_1),...,\Gamma_m^{+}(\tau_m)\}$. Note that the elements in $\Gamma(\tau)$ are in general correlated. 

Moreover, here we are particularly interested in the uniform inference across individual $\tau_j$'s because we need $m$-dimensional maps of level $\tau$ for given values $X=x$. Thus, from equation \eqref{eq:asympB}, we have that 
\begin{equation}
     \sqrt{n} \{ \hat{Q}(x,\tau) - Q(x,\tau) \}:=\sqrt{n} \{ \hat{B}(\tau)^\top x^\top - B(\tau)^\top x^\top \}
	\Rightarrow \Gamma(\tau)^\top x^\top.
\end{equation}

Consider now the effect of the generated regressors $\hat z_{ji}$. For notational purposes, in what follows we omit the quantile indexes $(\tau_1,...,\tau_m)$ in the $\beta$ coefficients whenever these are not explicitly  required. From Lemma \ref{thm1}, we define the influence functions for variable $j$ QR model as
\begin{equation*}
\psi_{ji}=\psi_j((y_{ji},z_{ji},),\beta_{j}^+)=\Omega_{1j}(\tau_j)^{-1}  \left [ \tau_j - I ( y_{ji} \leq z_{ji}\beta_{j}^+ ) \right ] z_{ji}^\top.
\end{equation*}
For convenience, we can write this function as
\begin{equation*}
\psi_{ji}=\Omega_{1j}(\tau_j)^{-1}  \left\{ \tau_j - I [ y_{ji} \leq z_{ji}(\beta_{-j}^+)\beta_{j}^+] \right\} z_{ji}(\beta_{-j}^+)^\top,
\end{equation*}
where we explicitly write the fact that the parameters 
\begin{equation*}
\beta_{-j}^+:=(\beta_{1}^+,...,\beta_{j-1}^+)
\end{equation*}
enter into the previous estimation of $z_{ji}$ (again note we omit the dependence on the quantile indexes). Note that this is a $k_{-j}\times 1$ vector where $k_{-j}=k\times k_2... \times k_{j-1}$.
Define $\psi_{ji}$ as the influence function evaluated at the true parameter values, and $\psi_{-ji}$ as the collection of the influence functions of the $-j$ parameters, also evaluated at the true $\beta$s.

The generated regressor involves indicator functions, which are themselves not differentiable. As far as we know, the issue of non-differentiable generated regressors has not been entirely solved in the literature. 
As an alternative, we follow the literature (see, e.g., \cite{Horowitz98}, \cite{KaplanSun17} and \cite{DeCastro2019}) and employ a smoothing mechanism, where we smooth the indicator function for the generated regressors. \cite{Horowitz98} shows that if the bandwidth converges to zero sufficiently rapidly, then the estimators based on the original indicator function and those of the smoothed version are asymptotically equivalent.\footnote{Another alternative is to implement it using estimating functions also with smoothing mechanisms as in \cite{KaplanSun17} and \cite{DeCastro2019}, and then to implement the sequential estimator as a generalized method of moments one. This would be based on the first order conditions derived from the QR set-up, and also on a smoothed version of the conditioning indicators for the generated regressors $z_{ji}$.} This smoothing technique applies to the conditioning dummy variables. We then consider generated regressor theory for recursive quantile regression as in \cite{MaKoenker06} and \cite{ChenGalvaoSong21}.

Let $\tilde I(\cdot)$ be a smoothing alternative for the $I(\cdot)$ indicator function with bandwidth $b_n$ (note we use the same function and bandwidth for all cases).
We then define  $\tilde d_{ji}=\tilde I(y_{ji}\leq \tilde z_{ji}\beta^+_j)=\tilde I(u_{ji}\leq 0)$, $j=1,2,...,m$ as a smoothed version of the conditioning indicators for the generated regressors $z_{ji}$, say $\tilde z_{ji}$. The latter would allow us to take derivatives with respect to the generated regressors.

\begin{assumption}\label{assump2}
The function $\tilde I(\cdot)$ satisfies $\tilde I(u)=0$  for $u \leq -1$, $\tilde I(u)=1$ for $u \geq 1$, and $-1 \leq \tilde I(u) \leq 2$ for $-1 < u < 1$.
The derivative $\tilde I'(.)$ is a symmetric, bounded kernel function of order $r \geq 2$, so $\int_{-1}^1\tilde I'(u)du=1$, $\int_{-1}^1u\tilde I'(u)du=0$ for
 $s =1,...,r-1$, and $\int_{-1}^1|u^r|\tilde I(u)du<\infty$ but $\int_{-1}^1u^r\tilde I(u)du\neq 0$, and 
$\int_{-1}^1|u^{r+1}|\tilde I(u)du<\infty$. The bandwidth sequence $b_n$ satisfies $b_n = o(n^{-1/(2r)})$.

For $j=2,...,m$ define (i) $\tilde z_j=\tilde g(w_j,\beta_{-j}^+)$, where $w_j=(y_{-j},x)$ contains all variables used in the construction of $z_j$ and using $\tilde I(\cdot)$ in place of the indicator function; (ii)  $z_j^*=g^*(w_j,\beta_{-j}^+)$ where we replace the corresponding distribution functions in Assumption \ref{assump1}-A2 in place of the indicator functions, that is, $I(y_{ji}\leq \tilde z_{ji}\beta^+_j)$ is replaced by $F_j(y_{ji}- z_{ji}\beta^+_j|z_{ji})$. Further define $G_{ij}=\nabla_{\beta_{-j}^+}  g({w}_{ji}, \beta_{-j}^+)^\top$ and $G_{ij}^*=\nabla_{\beta_{-j}^+}  g^*({w}_{ji}, \beta_{-j}^+)^\top$.
We assume that for any dominated function $\phi(w_j)$,
\begin{align}
\frac{1}{nb_n} \sum_{i = 1}^n \phi(w_{ji})G_{ji} \stackrel{p}{\rightarrow} E[\phi(w_{ji})G_{ji}^*], \ n\rightarrow\infty.
\end{align}
\end{assumption}





Then, the asymptotic distribution of the MQR estimators can be obtained as follows.

\begin{theorem}\label{thm2} Consider Assumptions \ref{assump1}-\ref{assump2}. Then, for $j=1,...,m$, the estimators $\tilde{\beta}_j^+$ are consistent and asymptotically normal, i.e., $\sqrt{n} ( \tilde{\beta}_j^+ - \beta_j^+) \stackrel{d}{\rightarrow}Normal(0_{k_j},V_j)\}$ as $n\rightarrow\infty$ where $V_j$ is the variance-covariance matrix that takes into account the generated regressor effect. 
\end{theorem}

The theorem above uses independent processes but it can be extended to time-series dependent data. In particular, \cite{GalvaoKatoMontesOlmo14}  has a similar representation to Lemma \ref{thm1} assuming that in Assumption \ref{assump1}-A1 $(y_{1i},y_{2i},...,y_{mi}.x_{i}), \ i=1,2,...\}_{i=1}^n$  is strict stationary and $\beta$-mixing with $\beta$-mixing coefficients satisfying $\beta(l) = O(l^{-\lambda})$ with $\lambda > \phi/(\phi-2)$, where $\phi > 2$ satisfies $E[\| (\bm{y}_{i},x_{i}) \|^{\phi}] < \infty$ for some $\phi > 2$. As such the Bahadur representation follows and consistency and asymptotic normality can be applied as in Theorem \ref{thm2}. However, the variance-covariance formula needs to be changed to dependent data.

Estimation of $V_j$ involves estimating  a convoluted formula based on the joint distribution of $Y_1,...,Y_j|X$. Similar to QR inference procedures, it requires  to compute densities, which are cumbersome for multivariate case. Then, bootstrap methods provide a useful alternative and whose validity is similar to that required for QR analysis.

\section{Comparison with other approaches}\label{comparison}

In \cite{HPS10} the multivariate model is decomposed into a series of univariate models. Quantiles are analyzed in terms of a \textit{magnitude} and a \textit{direction}. We define 
\(T = (\tau_1, \tau_2, \ldots, \tau_m) \in (0,1)^m\) as a set of quantile indices. 
The vector \(T\) can be factorized as \(T \equiv \tau v\), where 
\(\tau = \|T\| \in (0,1)\) represents the \textit{magnitude}, and 
\(v \in \mathcal{V}^{m-1} \equiv \{v \in \mathbb{R}^m : \|v\| = 1\}\) 
represents the \textit{direction}.

In this framework, \(\tau\) is a scalar quantile index that specifies the position along the distribution, 
while \(v\) is a unit vector that determines the direction in the \(m\)-dimensional space. 
This vector can be interpreted as an \((m - 1)\)-dimensional directional component that captures how quantile changes unfold across variables. This decomposition allows for an intuitive and geometric interpretation of multivariate quantiles in terms of distance and orientation within the variable space.

Vector Directional Quantiles (VDQ) are defined using the model
\begin{equation*}
Q_\tau(v^\top y \mid \Gamma_v^\perp y, x) 
= \Gamma_v^\perp y \gamma(v, \tau)
+ x\beta(v, \tau),
\end{equation*}
where \(\Gamma_v^\perp\) is an orthonormal basis of the subspace orthogonal to \(v\). 
This formulation captures the conditional quantile of the projection of \(y\) along the direction \(v\), 
given the orthogonal components of \(y\) and covariates \(x\). VDQ offers a powerful generalization 
of univariate quantiles, enabling the study of directional effects and distributional asymmetries in 
multivariate settings.

\begin{figure}
\centering
    \caption{Vector directional quantile}
    \label{fig:5}
    \includegraphics[width=8cm]{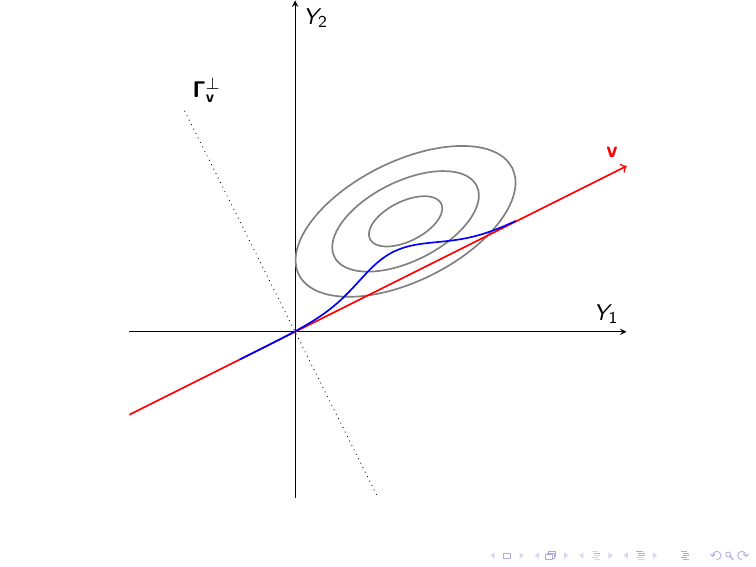}
\end{figure}

An alternative method is vector quantile regression (VQR) of \cite{ChernozhukovGalichonHallinHenry15} and \cite{CarlierChernozhukovGalichon16, CarlierChernozhukovGalichon17}. They develop a linear vector quantile regression framework that defines a quantile map $Q(X,T)$, where $T=(\tau_1,...,\tau_m)\in(0,1)^m$ as the gradient of a convex function, yielding a monotone mapping in the sense that $( Q(X,T) -  Q(X,T'))^\top (T - T') \geq 0$. 

Our proposed method differs from previous ones in several dimensions. First, it proposes a unique $\tau$-level QR coefficients interpretation, that is common in all vector directions. The quantile index corresponds to the joint multivariate distribution. This is of interest to researchers interested in the lower-tail realizations in the multivariate space. Second, it delivers a $m$-dimensional quantile graph. Third, asymptotic properties are based on univariate QR analysis with generated regressors.

\section{Monte Carlo}\label{sec:MC}

We consider data generating processes similar to the one used in Section \ref{bivariate}. Take the bivariate model for $(Y_1,Y_2)$:
\begin{align*}
    Y_1&=\beta_{10}+\beta_{11} X + (1+\alpha_{1}|X|)\varepsilon_{1},\\
    Y_2&=\beta_{20}+\beta_{21} X +     \gamma_{21} Y_1+ (1+\alpha_{2}|X|)\varepsilon_{2}.   
\end{align*}
Here we assume that $X$, $\varepsilon_{1}$ and $\varepsilon_{2}$ are independent $\mathcal{N}(0,1)$, $\beta_{10}=\beta_{11}=\beta_{20}=\beta_{21}=\gamma_{21}=1$. Then, conditional on $X$ we obtain a bivariate normal distribution of $(Y_1,Y_2)$.

We consider two cases, one for the location shift model, $\alpha_{1}=\alpha_{2}=0$, in Table \ref{tab:a0}, and one for the location-scale shift model, $\alpha_{1}=\alpha_{2}=1$, in Table \ref{tab:a1}. We evaluate the experiments for the conditional model of $X=1$ and for both orders $1\rightarrow2$ and $2\rightarrow1$. Moreover, we examine the quantile models of $\tau\in\{0.25,0.50,0.75\}$ and $n\in\{100,200,500,1000\}$. For each case we consider 1000 Monte Carlo replications.

The performance of the estimators is analyzed in terms of the estimated quantile graph. In particular, for each element of the estimated graph $(\hat q_1(x=1),\hat q_2(x=1))$, say $h=1,...,H$, we compute the average probability $\hat p=\frac 1 H \sum_{h=1}^H P[(Y_1\leq \hat q_1^\tau(X=1,h))\&(Y_2\leq \hat q_2^\tau(X=1,h))]$ and compare it with the theoretical $\tau$. Then we report bias, variance, and mean squared error (MSE) of $\hat p$.

\begin{table}
\centering
\caption{Results for $\alpha_{1}=\alpha_{2}=0$}
\label{tab:a0}
\begin{tabular}{rllllllll}
\toprule
   $n$ & $\tau$ & bias $1 \rightarrow 2$ & var $1 \rightarrow 2$ & mse $1 \rightarrow 2$ & bias $2 \rightarrow 1$ & var $2 \rightarrow 1$ & mse $1 \rightarrow 2$ \\
\midrule
 100 & 0.25 &   -0.0180 &   0.0039 &   0.0042 &    0.0226 &   0.0035 &     0.0040 \\
 200 & 0.25 &   -0.0213 &   0.0021 &   0.0026 &    0.0208 &   0.0018 &     0.0023 \\
 500 & 0.25 &   -0.0217 &   0.0008 &   0.0013 &    0.0200 &   0.0007 &     0.0011 \\
1000 & 0.25 &   -0.0229 &   0.0004 &   0.0009 &    0.0196 &   0.0003 &     0.0007 \\
 100 & 0.50 &   -0.0299 &   0.0059 &   0.0068 &    0.0109 &   0.0047 &     0.0049 \\
 200 & 0.50 &   -0.0301 &   0.0031 &   0.0040 &    0.0118 &   0.0024 &     0.0026 \\
 500 & 0.50 &   -0.0267 &   0.0012 &   0.0019 &    0.0145 &   0.0009 &     0.0011 \\
1000 & 0.50 &   -0.0259 &   0.0006 &   0.0013 &    0.0152 &   0.0005 &     0.0007 \\
 100 & 0.75 &   -0.0310 &   0.0050 &   0.0060 &   -0.0054 &   0.0036 &     0.0037 \\
 200 & 0.75 &   -0.0221 &   0.0025 &   0.0030 &   -0.0008 &   0.0018 &     0.0018 \\
 500 & 0.75 &   -0.0173 &   0.0009 &   0.0012 &    0.0040 &   0.0007 &     0.0007 \\
1000 & 0.75 &   -0.0149 &   0.0005 &   0.0007 &    0.0058 &   0.0004 &     0.0004 \\
\bottomrule
\end{tabular}
\end{table}

\begin{table}
\centering
\caption{Results for $\alpha_{1}=\alpha_{2}=1$}
\label{tab:a1}
\begin{tabular}{rllllllll}
\toprule
   $n$ & $\tau$ & bias $1 \rightarrow 2$ & var $1 \rightarrow 2$ & mse $1 \rightarrow 2$ & bias $2 \rightarrow 1$ & var $2 \rightarrow 1$ & mse $1 \rightarrow 2$ \\
\midrule
 100 & 0.25 &    0.0103 &   0.0040 &   0.0041 &    0.0380 &   0.0037 &     0.0051 \\
 200 & 0.25 &    0.0070 &   0.0021 &   0.0022 &    0.0364 &   0.0020 &     0.0033 \\
 500 & 0.25 &    0.0065 &   0.0008 &   0.0008 &    0.0359 &   0.0007 &     0.0020 \\
1000 & 0.25 &    0.0054 &   0.0004 &   0.0004 &    0.0354 &   0.0004 &     0.0016 \\
 100 & 0.50 &   -0.0600 &   0.0052 &   0.0088 &   -0.0288 &   0.0045 &     0.0053 \\
 200 & 0.50 &   -0.0606 &   0.0027 &   0.0063 &   -0.0284 &   0.0023 &     0.0031 \\
 500 & 0.50 &   -0.0574 &   0.0010 &   0.0042 &   -0.0261 &   0.0008 &     0.0015 \\
1000 & 0.50 &   -0.0579 &   0.0006 &   0.0039 &   -0.0257 &   0.0004 &     0.0011 \\
 100 & 0.75 &   -0.1019 &   0.0059 &   0.0163 &   -0.0724 &   0.0045 &     0.0097 \\
 200 & 0.75 &   -0.0934 &   0.0029 &   0.0116 &   -0.0681 &   0.0023 &     0.0069 \\
 500 & 0.75 &   -0.0901 &   0.0011 &   0.0092 &   -0.0640 &   0.0009 &     0.0050 \\
1000 & 0.75 &   -0.0875 &   0.0006 &   0.0083 &   -0.0627 &   0.0005 &     0.0044 \\
\bottomrule
\end{tabular}
\end{table}

The results show that the performance of the estimators improves as the sample size \(n\) increases, with reductions in both variance and MSE. For both settings of \(\alpha_1 = \alpha_2 = 0\) and \(\alpha_1 = \alpha_2 = 1\), bias is generally small and decreases as the sample size increases. However, bias is more pronounced in the case of \(\alpha_1 = \alpha_2 = 1\), especially for smaller sample sizes, indicating a more complex structure that affects the estimator's performance. Variance is consistently reduced with larger sample sizes, and MSE also shows a clear pattern of reduction with increasing \(n\). Moreover, higher quantiles, such as \(\tau = 0.75\), show higher MSE, particularly for smaller sample sizes, highlighting the difficulty in accurately estimating higher quantiles with limited data (i.e., the larger $\tau$ is, there will be less quantile indexes to be used for the first variable in the ordering).

In general, estimators perform better at smaller quantiles (\(\tau = 0.25\) and \(\tau = 0.50\)), where both bias and MSE are lower compared to larger quantiles. For the case \(\alpha_1 = \alpha_2 = 0\), the estimators exhibit minimal bias and MSE across sample sizes, with variance decreasing steadily as \(n\) increases. In contrast, for \(\alpha_1 = \alpha_2 = 1\), the bias and MSE are higher, especially for smaller sample sizes and larger quantiles, reflecting the added complexity introduced by non-zero \(\alpha\) parameters.

\section{Empirical application: Pass-through in Argentina 2004-2024} \label{application}

\subsection{Pass-through for dollarized economies}

The coefficient or elasticity of exchange rate pass-through (ERPT) to domestic prices—hereafter referred to as pass-through—measures the extent to which changes in the exchange rate (ER) are transmitted to domestic prices. In the case of Argentina, a high-inflation economy where the U.S. dollar serves as the primary currency of reference (i.e., a dollarized or bimonetary economy), understanding the magnitude and persistence of ERPT is central to macroeconomic analysis. Beyond its effect on prices, ERPT is also critical due to its implications for output. Periods of economic expansion tend to generate increased demand for foreign currency, whether for imports or for saving in foreign-denominated assets. This demand imposes a binding external constraint—a phenomenon commonly referred to as stop-and-go dynamics.

Pass-through effects are traditionally estimated using vector autoregression (VAR) models with ordinary least squares (OLS), accompanied by impulse response function (IRF) analysis based on conditional means. However, such approaches are limited in their ability to capture asymmetric and heterogeneous dynamic responses, which are frequently observed in ERPT, particularly in emerging and developing economies. Nonlinear models are combined with VAR-OLS analysis.

A well-documented feature of ERPT in these economies is the presence of substantial asymmetries and heterogeneity—both across countries and across different time periods within the same country. These characteristics underscore the need for more flexible empirical strategies that can account for variation in the transmission of ER shocks.

We consider monthly data from Argentina, January 2004 to December 2024. Constructing consistent macroeconomic series for Argentina, particularly during the 2007–2015 period, poses a well-known challenge due to issues with official statistics.

First, during those years, the official statistical agency--INDEC (Instituto Nacional de Estadísticas y Censos)--was widely regarded as having manipulated inflation data to under report consumer price increases. Consequently, the consumer price index (CPI) must be constructed from alternative sources. Indeed, the IMF does not report Argentine CPI for this period. The procedure we adopt is as follows: we use INDEC data for January 2004 to December 2006, the CPI from the Ciudad Autónoma de Buenos Aires for January 2007 to May 2016, and again INDEC data from June 2016 to December 2018.

Second, Argentina implemented ER controls during two periods: 2011–2015 and 2019–2023. In both cases, the relevant rate was the official exchange rate applied to imports and exports. We use the monthly average of this rate as reported by the Central Bank of Argentina, available at \url{http://www.bcra.gov.ar/pdfs/operaciones/com3500.xls}.

Third, for output, we rely on the industrial production index published by INDEC—specifically, the Estimador Mensual de Actividad Económica (EMAE), which is seasonally adjusted. Importantly, this series is not considered to have been manipulated during the 2007–2015 period.

A stylized narrative of Argentine macroeconomic dynamics over this period is as follows. Between 2004 and 2009, the economy experienced sustained high growth and moderate inflation. From 2007 onward, inflation began to rise, closely tracking movements in the exchange rate. The exchange rate was actively used as a policy tool to anchor prices; periods of stability were often followed by sharp devaluations, typically triggered by speculative attacks on the peso. After a period of strong growth (2004–2011), output stagnated in 2012. Argentina entered a debt crisis in May 2018, which was soon compounded by the COVID-19 recession. Since 2018, the country has experienced persistent high inflation and continued currency depreciation.

\begin{figure}[ht]
    \centering
       \caption{Log First Differences of CPI, EMAE, and official ER}
        \label{fig:series}        \includegraphics[width=0.8\textwidth]{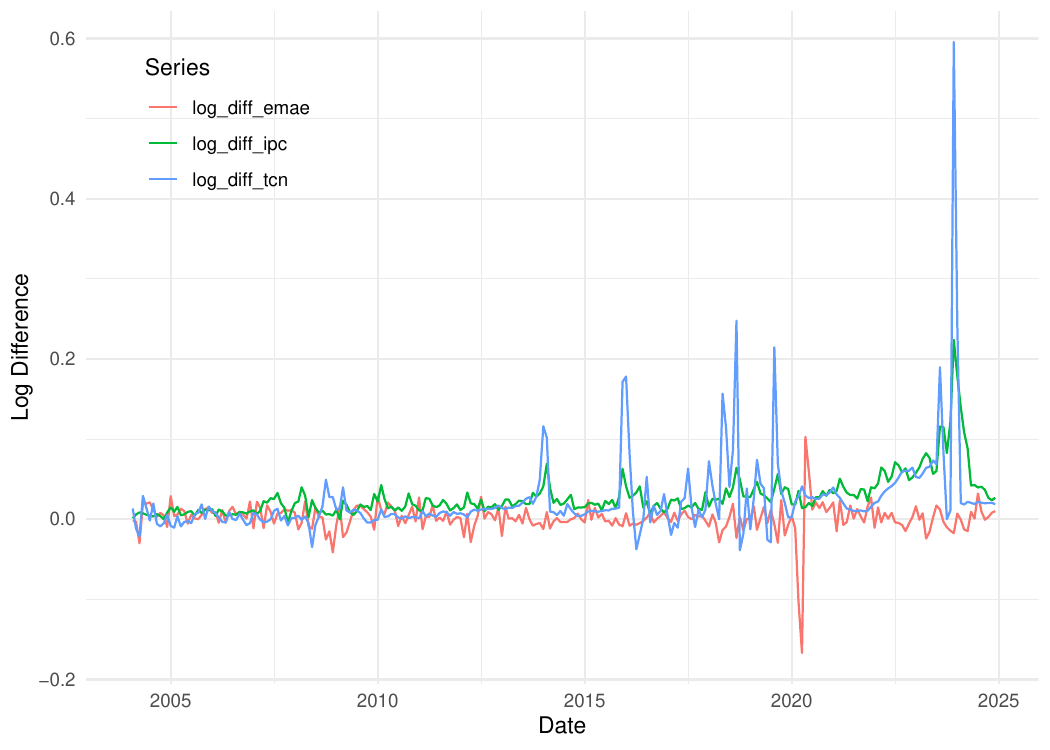}
\end{figure}

Figure \ref{fig:series} plots the log-difference series of the key variables.

\subsection{Econometric analysis}\label{results}

For further reference, let  $y$ be the output growth measure, without seasonality, $p$ be the inflation rate calculated as the monthly first difference of the logarithm of the CPI, $e$ monthly first difference of the logarithm of the exchange rate.  In all cases, unit-root tests applied to these transformed series are rejected and thus they are stationary. Using the Bayesian information criterion for the VAR-OLS model for the three variables we use a model with one lag. 

The typical procedure used in VAR models to identify shocks is to impose an ordering for the Cholesky decomposition. Output $y$ is assumed to have a contemporaneous effect on the other variables in the system, $p$ and $e$, while the opposite does not occur. $p$ contemporaneously affects $e$ but not $y$. Finally, $e$ has no contemporaneous effects on $y$ and $p$. This model implies that the exchange rate is the nominal variable that adjusts first, then prices, and finally output. This ordering is consistent with the effect of devaluations in Argentina: when there is an exchange rate devaluation, prices and output take some time to adjust. Moreover, the official ER is managed by the government to some extent.
The MQR model is estimated for a bivariate model
 $\tau$ using $(\tau_y,\tau_p)\in (0,1)^2$  that satisfy the sequential conditioning restriction with a set of control variables $x_t=(e_t,y_{t-1},p_{t-1},e_{t-1})$. 
 
Consider the order $y\rightarrow p$. For this case, $\tau_y\in(\tau,1)$ and $\tau_p=\tau/\tau_y$ and the QR models are:
\begin{align*}
q_y^{y\rightarrow p}(x_t) &= Q_{\tau_y}\{y_t|e_t,(y_{t-1},p_{t-1},e_{t-1})\} = \\
  &\quad \beta_{y0}(\tau_y) + \beta_{ye}(\tau_y)e_t + \beta_{yy-1}(\tau_y)y_{t-1} + \beta_{yp-1}(\tau_y)p_{t-1} + \beta_{ye-1}(\tau_y)e_{t-1}, \\
q_p^{y\rightarrow p}(x_t) &= Q_{\tau_p}\{p_t|e_t,(y_{t-1},p_{t-1},e_{t-1}),y_t\leq q_y^{y\rightarrow p}(x_t)\} = \\
  &\quad I[y_t \leq q^{y\rightarrow p}(x_t)]\times \\
  &\quad \left( \beta_{p0}(\tau_p) + \beta_{pe}(\tau_p)e_t + \beta_{py-1}(\tau_p)y_{t-1} + \beta_{pp-1}(\tau_p)p_{t-1} + \beta_{pe-1}(\tau_p)e_{t-1} \right), \\
  &\quad + (1 - I[y_t \leq q^{y\rightarrow p}(x_t)])\times \\
 &\quad  \left( \beta^-_{p0}(\tau_p) + \beta^-_{pe}(\tau_p)e_t + \beta^-_{py-1}(\tau_p)y_{t-1} + \beta^-_{pp-1}(\tau_p)p_{t-1} + \beta^-_{pe-1}(\tau_p)e_{t-1} \right).
\end{align*}

For this case, $\tau_p\in(\tau,1)$ and $\tau_y=\tau/\tau_p$, and the QR models are:
\begin{align*}
q_p^{p\rightarrow y}(x_t) &= Q_{\tau_p}\{p_t|e_t,(y_{t-1},p_{t-1},e_{t-1})\} = \\
  &\quad \beta_{p0}(\tau_p) + \beta_{pe}(\tau_p)e_t + \beta_{py-1}(\tau_p)y_{t-1} + \beta_{pp-1}(\tau_p)p_{t-1} + \beta_{pe-1}(\tau_p)e_{t-1}, \\
q_y^{p\rightarrow y}(x_t) &= Q_{\tau_y}\{y_t|e_t,(y_{t-1},p_{t-1},e_{t-1}),p_t\leq q_p^{p\rightarrow y}(x_t)\} = \\
  &\quad I[p_t \leq q^{p\rightarrow y}(x_t)]\times \\
  &\quad \left( \beta_{y0}(\tau_y) + \beta_{ye}(\tau_y)e_t + \beta_{yy-1}(\tau_y)y_{t-1} + \beta_{yp-1}(\tau_y)p_{t-1} + \beta_{ye-1}(\tau_y)e_{t-1} \right), \\
  &\quad + (1 - I[p_t \leq q^{p\rightarrow y}(x_t)])\times \\
 &\quad  \left( \beta^-_{y0}(\tau_y) + \beta^-_{ye}(\tau_y)e_t + \beta^-_{yy-1}(\tau_y)y_{t-1} + \beta^-_{yp-1}(\tau_y)p_{t-1} + \beta^-_{ye-1}(\tau_y)e_{t-1} \right).
\end{align*}
Note that in these models, $\beta^+=(\beta^\top,\beta^{-\top})^\top$.

We estimate the effect of ERPT within the same month as the devaluation by considering two distinct magnitudes of depreciation: $\Delta r = 0.1$ and $\Delta r = 0.2$, corresponding to 10\% and 20\% currency depreciation, respectively. The lagged values of the control variables are fixed at their historical averages to isolate the contemporaneous effect of the shock.

Our analysis focuses first on the joint distribution of the variables $(y, p)$ and $(y, -p)$, where $y$ denotes output growth and $p$ represents inflation. The $(y, p)$ specification captures scenarios with lower output growth and/or lower inflation, while the $(y, -p)$ configuration highlights cases of lower output growth and/or higher inflation, which are particularly relevant for policy evaluation. In this alternative specification, $-p$ is interpreted as deflation and thus has the opposite sign of inflation—more negative values of $-p$ imply higher inflation. Consequently, in the $(y, -p)$ space, worse inflation scenarios appear further to the left on the horizontal axis.

To estimate the conditional quantile response, we compute the marginal models over a fine quantile grid with a spacing of 0.01. For visualization and interpretation, we fit a linear regression model of $q_2$ on $q_1$ and $1/q_1$.

\begin{figure}[!htbp]
    \centering
    \caption{$\tau=0.25$ effects of exchange rate devaluation on output growth and inflation, $\Delta r=\{0.1,0.2\}$.}
    \label{fig:Dev}
    \begin{subfigure}[b]{0.6\textwidth}
        \centering
        \includegraphics[width=\textwidth]{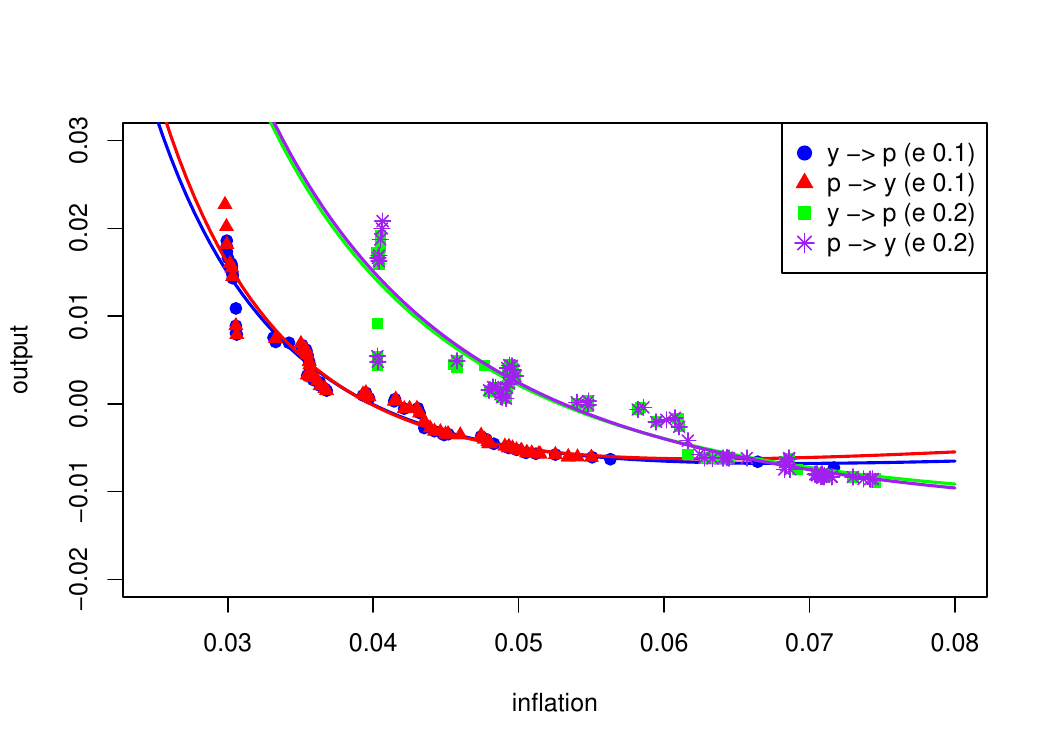}
        \caption{Inflation}
        \label{fig:Dev_a}
    \end{subfigure}
    \vfill
    \begin{subfigure}[b]{0.6\textwidth}
        \centering
        \includegraphics[width=\textwidth]{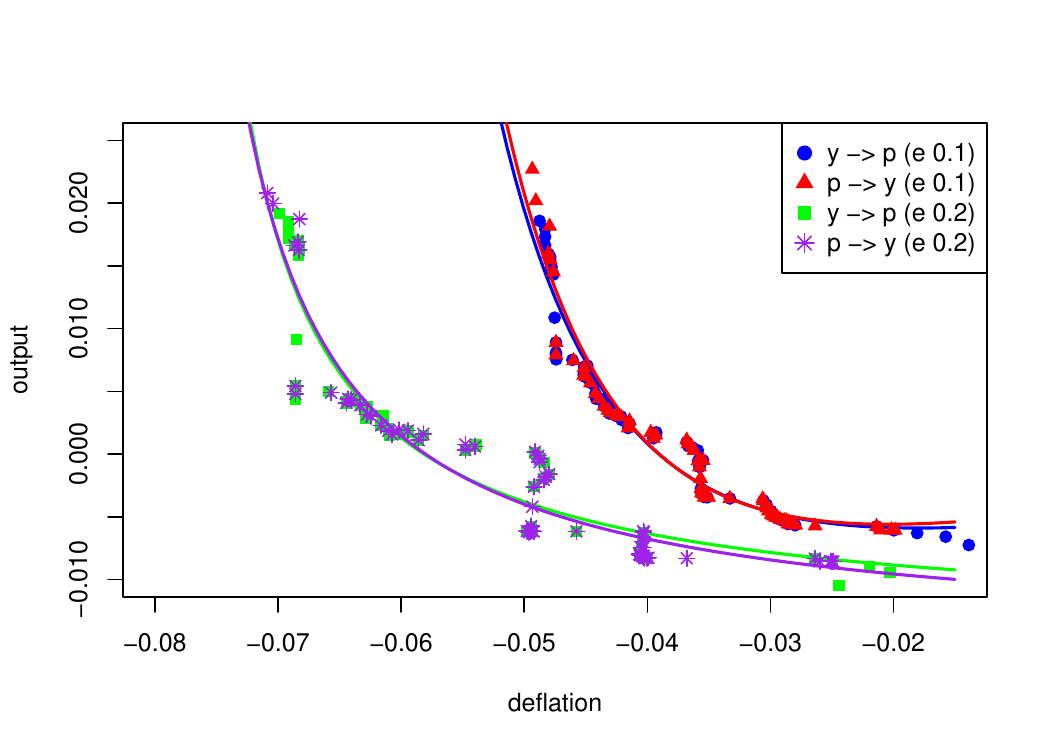}
        \caption{Deflation (=(-1)*Inflation)}
        \label{fig:Dev_b}
    \end{subfigure}

    \caption*{Notes: Fitted lines correspond to a linear regression of $q_2$ on $q_1$ and $1/q_1$.}

\end{figure}

Figures \ref{fig:Dev_a} and \ref{fig:Dev_b} display the estimated quantile graphs for a fixed quantile $\tau=0.25$. In both cases, there is a clear impact of an ER devaluation: it increases the likelihood of high-inflation scenarios. Moreover, larger devaluations are associated with a worsening of output-inflation trade-offs, leading to more adverse combinations of lower output and higher inflation (or reduced deflation). Both figures illustrate the asymmetric impact of an ER devaluation. Notably, the rightward shift in the second graph is approximately twice as large as in the first. This suggests that adverse inflation scenarios are more strongly affected by a devaluation than favorable ones. In other words, the likelihood and intensity of high-inflation outcomes increase disproportionately, highlighting the nonlinear and asymmetric effects of ER shocks on inflation dynamics.

\begin{figure}[!htbp]
    \centering
    \caption{$\tau=\{0.10,0.20,0.30,0.40,0.50\}$ effects of exchange rate devaluation on output growth and inflation, $ r=\{0.1,0.2\}$.}
    \label{fig:Dev_all}
    \begin{subfigure}[b]{0.6\textwidth}
        \centering
        \includegraphics[width=\textwidth]{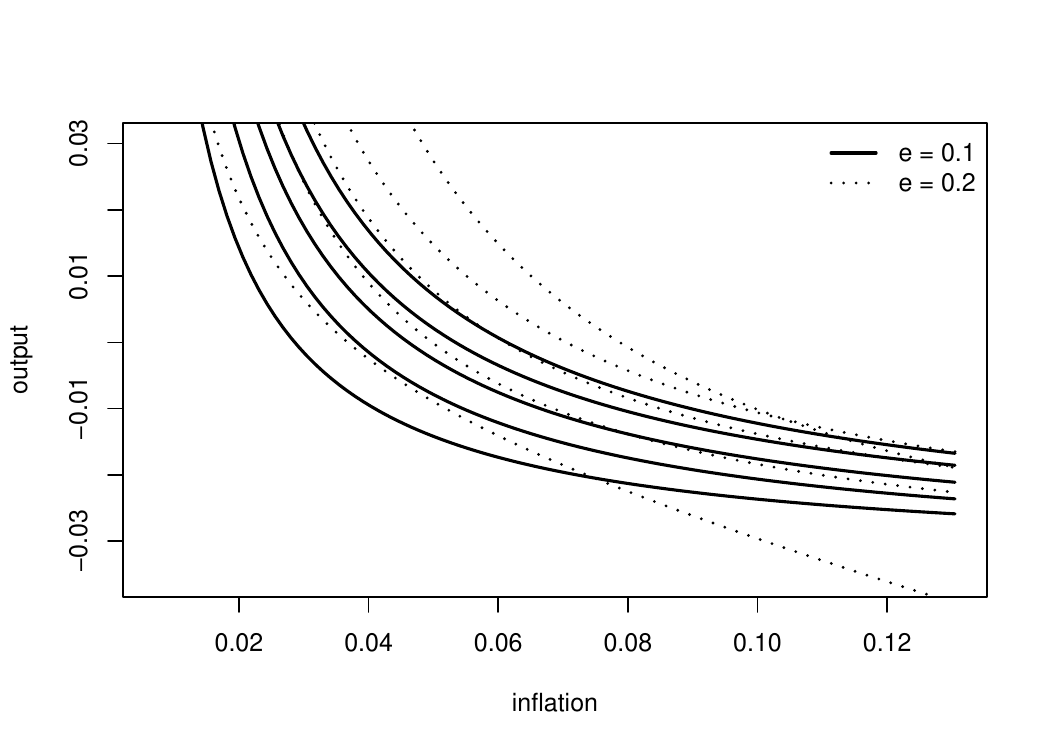}
        \caption{Inflation}
        \label{fig:Dev_all_a}
    \end{subfigure}
    \vfill
    \begin{subfigure}[b]{0.6\textwidth}
        \centering
        \includegraphics[width=\textwidth]{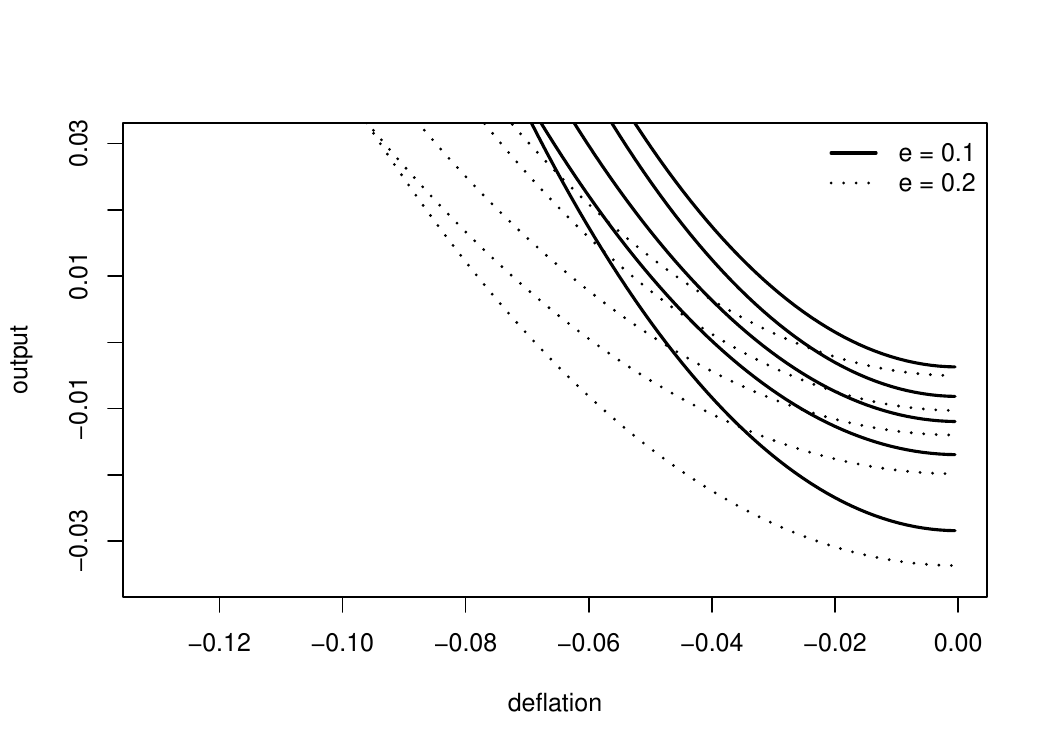}
        \caption{Deflation (=(-1)*Inflation)}
        \label{fig:Dev_all_b}
    \end{subfigure}
   \caption*{Notes: Fitted lines correspond to a linear regression of $q_2$ on $q_1$ and $1/q_1$.}
\end{figure}

Next we consider the joint evaluation of several quantiles. Figure \ref{fig:Dev_all} consider the cases of $\tau\in\{0.1,0.2,0.3,0.4,0.5\}$ for the output-inflation and output-deflation bivariate cases. Here we report the $y\rightarrow p$ order only, similar results are obtained with the other order. The figures illustrate how the devaluation impacts on the joint distribution and thus how it shifts the entire distribution.

\section{Conclusion}\label{conclusion}
This paper introduces a new framework for modeling multivariate quantiles based on the multivariate distribution function, termed multivariate quantile regression (MQR). Unlike existing methods such as directional quantiles or vector quantile regression, MQR relies on the conditional probability structure of joint distributions to define and estimate quantile curves. By leveraging a sequence of univariate quantile regressions conditioned on subsets of the outcome space, the model offers a flexible and interpretable approach to studying the joint distributional behavior of multiple endogenous variables in response to observed covariates.

The theoretical contribution of the paper lies in the formal development of the MQR estimator under a sequential conditioning structure. These results highlight the consistency and asymptotic normality of the estimator under standard regularity conditions. Furthermore, we show that multiple permutations of the endogenous variables can be used to obtain alternative but equivalent representations of the multivariate quantile surface, analogous to identification schemes in structural models.

Through simulation exercises, we illustrate the finite-sample properties of the estimator under different model specifications, including location and location-scale shift models. The simulation results confirm the robustness of MQR across different data-generating processes and show that the permutation order may influence the precision of the estimates in finite samples. Importantly, both directions of estimation yield qualitatively similar quantile surfaces, reaffirming the estimator’s validity. 

The empirical application focuses on exchange rate pass-through in Argentina, a relevant and complex case given the country’s history of high inflation, dollarization, and external constraints. Using monthly data from 2004 to 2024, we apply MQR to evaluate the joint response of inflation and output to changes in the exchange rate. Our results reveal strong asymmetric effects: adverse inflation scenarios are more responsive to devaluation shocks than favorable ones. This highlights the nonlinear nature of pass-through and the limitations of relying solely on mean-based or symmetric VAR models in macroeconomic analysis for emerging markets.

In sum, the MQR approach provides a powerful tool for studying joint distributional responses in a multivariate setting. It opens the door to more nuanced policy analysis, particularly in environments where tail risks and asymmetric responses are central to economic dynamics. Future research may extend this framework to high-dimensional settings, dynamic panel structures, or time-varying quantile systems. In addition, further empirical applications—such as monetary policy evaluation, financial risk management, or inequality analysis—could benefit from the methodological flexibility and interpretability of the MQR model.

\newpage

\section*{Appendix}

\subsection*{Proof of Lemma \ref{thm1}}

\begin{proof}
The result follows from an application of the uniform convergence derivations for the univariate quantile regression case in \cite{GalvaoKatoMontesOlmo14}. In particular, note that the conditions in Assumption \ref{assump1} (A1)--(A4)  correspond to conditions (C1)--(C6) in \cite{GalvaoKatoMontesOlmo14} for a linear model.  Therefore, we can apply Lemma 1 and Theorem 1 of that paper, which correspond to the present lemma.
\end{proof}

\subsection*{Proof of Theorem \ref{thm2}}

\begin{proof}
For each estimator $j$, by Lemma \ref{thm1}, the following Bahadur representation holds
	\begin{equation*}
	\sqrt{n} \{ \hat{\beta}_j^+(\tau_j) - \beta_j^+(\tau_j) \} =  \Omega_{1j}(\tau_j)^{-1} \frac{1}{\sqrt{n}} \sum_{i=1}^n  \left [ \tau_j - I ( y_{ji} \leq z_{ji}\beta_{j}^+(\tau_j)  ) \right ] z_{ji}^\top
	+ o_p(1).
	\end{equation*}
Next, we need to adjust the Bahadur representation to consider the effect of the previous estimation of the QR parameters.

We follow the procedure in \cite{ChenGalvaoSong21} for generated regressors models in QR models, in particular Theorems  3.1 (consistency) and 3.2 (asymptotic normality). In particular, for each $j$, we take into account the $(m-j)$ QR coefficients that were previously estimated, thus giving a sequential structure. Note that all these coefficients have a similar Bahadur representation, thus a valid known influence function.

The proof follows these two steps.

Step 1. If the generated regressors were differentiable functions, the asymtptotic properties are obtained by applying \cite{MaKoenker06}, Lemma 2 G-inverse method for two-step recursive QR models, as in \cite{ChenGalvaoSong21}.

Step 2. These results require, however, that the generated regressor, i.e. $\hat z_j$, to be differentiable with respect to $\beta_{-j}^+$, which is not the case. However, using \cite{DeCastro2019} smoothing mechanism as in Assumption \ref{assump2} allows for taking derivatives of $\tilde z_j$ modified regressors, i.e. $\nabla_{\beta_{-j}^+}\tilde z_j(\beta_{-j}^+)$. This derivative includes a negligible contribution of the smoothing function with respect to the true regressor using the indicator function. This, however, given Assumption \ref{assump2} restriction on the bandwidth sequence, does not affect the asymptotic representation. 

\vspace{.5cm}

\noindent\textbf{Step 1. Assuming $\hat z_j(\beta_{-j}^+)$ is differentiable}

To simplify notation, in what follows we omit the use of the indexes $j$ and $-j$ unless required, we use just $\beta$ to refer to $\beta_j^+$ and $\theta$ to refer to $\beta_{-j}^+$. Moreover, we assume that $ z_j=g(w_j,\beta_{-j}^+)$, where $w_j=(y_{-j},x)$ contains all variables used in the construction of $z_j$. Finally, we will use $z_h=g(w_h,\theta_h)$, $h=1,...,k_j$ to refer to individual elements of $z_j$, such that $z_j=(z_1,...,z_{k_j})$. The following lines correspond to $j=2,...,m$ because for the case $j=1$ there is no previously generated regressor, and thus $z_1=x$.

Let $\check{{\beta}}$ be the two-step QR with generated regressors $\hat z$ estimator, $\hat{{\beta}}$ be the usual (infeasible) QR estimator with true unobservable variables, and $\beta$ the true parameter. Consider
\begin{align}
\label{eq:betahat}
\check{{\beta}} - {{\beta}}
=(\check{{\beta}} - \hat{{\beta}} ) + (\hat{{\beta}}  - {{\beta}}).
\end{align}

Under the standard QR conditions in Assumption \ref{assump1}, from Lemma \ref{thm1} we have 
\begin{equation}
 \hat{{\beta}} - {\beta} \overset{p}{\to} 0, \label{eq:second}
\end{equation}
and the following linear representation
\begin{equation}
\label{important1}
\hat{{\beta}}  - {{\beta}} = D_1^{-1} \frac{1}{n} \sum_{i = 1}^n \psi(y_i - {z}_i {\beta}) z_i^\top + o_p(1),
\end{equation}
where $\psi$ is the function $\psi(u) := \tau_j - I(u \le 0)$ (here we do refer to the $j$ case and use $\tau_j$) and $D_1=\Omega_{1j}(\tau_j)$.

Now, by noticing that $\check{{\beta}} = \check{{\beta}} (\hat{z}_1,\cdots, \hat{z}_q)$ and $\hat{{\beta}} = \hat{{\beta}} (z_1,\cdots, z_{q})$, and by expanding the first term in equation \eqref{eq:betahat} $\check{{\beta}} - \hat{{\beta}}$, we have
\begin{align*}
\check{{\beta}}(\tau) - \hat{{\beta}} (\tau) &=  \left(\frac{\partial \hat{{\beta}} (\tau)}{\partial z_1}^\top, \cdots, \frac{\partial \hat{{\beta}} (\tau)}{\partial z_q}^\top \right) \begin{bmatrix}
\hat{z}_1 - z_1 \\
\vdots \\
\hat{z}_q - z_q
\end{bmatrix} + o_p(1) \\
&=  \left(\frac{\partial \hat{{\beta}} (\tau)}{\partial z_1}^\top, \cdots, \frac{\partial \hat{{\beta}} (\tau)}{\partial z_q}^\top \right) \begin{bmatrix}
g_1({w}_1, \check{{\theta}}_1) - g_1({w}_1, {\theta}_1) \\
\vdots \\
g_q({w}_q, \check{{\theta}}_q) - g_q({w}_q, {\theta}_q)
\end{bmatrix} + o_p(1) \\
&= \left(\frac{\partial \hat{{\beta}} (\tau)}{\partial z_1}^\top, \cdots, \frac{\partial \hat{{\beta}} (\tau)}{\partial z_q}^\top \right) \begin{bmatrix}
\nabla_{\theta_1} g_1({w}_1, {\theta_1})^\top(\check{{\theta}}_1 - {\theta}_1) \\
\vdots \\
\nabla_{\theta_q} g_q({w}_q, {\theta_q})^\top(\check{{\theta}}_q - {\theta}_q)
\end{bmatrix} + o_p(1),
\end{align*}
where the last equality follows from an expansion for $g_h(\cdot)$ for $h = 1, \cdots, q$.

Now we apply Lemma 2 about G-inverse from \cite{MaKoenker06}, and obtain that for $h = 1, \cdots, q$
\begin{align*}
\left(\frac{\partial \hat{{\beta}} (\tau)}{\partial z_h}\right)^\top \nabla_{\theta_h} g_h({w}_h, {\theta}_h)^\top &= - D_1^{-1} \left(\frac{1}{n} \sum_{i=1}^n f(0 | {z}_i) \beta_h {z}_i^\top \nabla_{\theta_h} g_h({w}_{hi}, {\theta}_h)^\top \right) + o_p(1) \\
&= - D_1^{-1} D_{12}^h + o_p(1),
\end{align*}
where
\begin{equation}
D_{12}^h = \mbox{plim}_{n \to \infty} \frac{1}{n} \sum_{i = 1}^n f(0 | {z}_i)\beta_h {z}_i^\top\nabla_{\theta_h}  g_h({w}_{hi}, {\theta}_h)^\top.
\end{equation}
Here $f(0 | {z}_i)$ is the density function for the $j$ case as given in Assumption \ref{assump1}-(A2).

It follows that
\begin{align*}
\check{{\beta}} (\tau) - \hat{{\beta}} (\tau) &= \sum_{h=1}^q \left( \frac{\partial \hat{{\beta}} (\tau)}{\partial z_h} \right)^\top \nabla_{\theta_h}  g_h({w}_h, {\theta}_h)^\top(\check{{\theta}}_h - {\theta}_h) + o_p(1) \\
&= \sum_{h=1}^q (- D_1^{-1} D_{12}^h + o_p(1) ) (\check{{\theta}}_h - {\theta}_h) + o_p(1)\\
&= \sum_{h=1}^q [- D_1^{-1}D_{12}^h (\check{{\theta}}_h - {\theta}_h)] + o_p(1).
\end{align*}
For any $h = 1, \cdots, q$: $ \check{{\theta}}_h = {\theta}_h+o_p(1)$  and $D_1^{-1}D_{12}^h$ is bounded by Assumption \ref{assump1}, we have that $-D_1^{-1} D_{12}^h (\check{{\theta}}_h - {\theta}_h ) = o_p(1)$.
Therefore, as $n \to \infty$, we have
\begin{equation}
\check{{\beta}} - \hat{{\beta}}  \overset{p}{\to} 0. \label{eq:first} 
\end{equation}

Combining \eqref{eq:second} and  \eqref{eq:first}, we have
\begin{align*}
\check{{\beta}}  - {\beta} &= ( \check{{\beta}}  - \hat{{\beta}}  ) + ( \hat{{\beta}}  - \check{{\beta}} ) \overset{p}{\to} 0.
\end{align*}

Consider now the following
\begin{align}
\label{eq:n_betahat}
\sqrt{n}(\check{{\beta}}(\tau) - {\beta_0} (\tau))
=\sqrt{n}(\check{{\beta}}(\tau) - \hat{{\beta}}(\tau)) + \sqrt{n}(\hat{{\beta}}(\tau) - {\beta}_0(\tau)).
\end{align}

Under the QR conditions imposed in Assumption \ref{assump1}, we have that the second term in equation \eqref{eq:n_betahat} has the standard QR expansion as
\begin{equation}
\label{eq:Bahadur_rep}
\sqrt{n}(\hat{{\beta}} - {\beta} ) = D_1^{-1} \frac{1}{\sqrt{n}} \sum_{i = 1}^n  \psi(y_i - {z}_i {\beta}) {z}_i^\top+ o_p(1),
\end{equation}
and satisfies a Central Limit Theorem such that
\begin{equation}
\label{eq:n_second} 
\sqrt{n}(\hat{{\beta}} - {\beta})  \overset{d}{\to}  N(0, \tau_j(1 - \tau_j)D_1^{-1}D_0D_1^{-1}),
\end{equation}
where we define $D_0=\Omega_{0j}$.

Now, for the second term,
\begin{align*}
\sqrt{n} (\check{{\beta}}(\tau) - \hat{{\beta}} (\tau)) &=  \sqrt{n}\left(\frac{\partial \hat{{\beta}} (\tau)}{\partial z_1}^\top, \cdots, \frac{\partial \hat{{\beta}} (\tau)}{\partial z_q}^\top \right) \begin{bmatrix}
\hat{z}_1 - z_1 \\
\vdots \\
\hat{z}_q - z_q
\end{bmatrix} + o_p(1) \\
&= \sqrt{n} \left(\frac{\partial \hat{{\beta}} (\tau)}{\partial z_1}^\top, \cdots, \frac{\partial \hat{{\beta}} (\tau)}{\partial z_q}^\top \right) \begin{bmatrix}
g_1({w}_1, \check{{\theta}}_1) - g_1({w}_1, {\theta}_1) \\
\vdots \\
g_q({w}_q, \check{{\theta}}_q) - g_q({w}_q, {\theta}_q)
\end{bmatrix} + o_p(1) \\
&= \sqrt{n} \left(\frac{\partial \hat{{\beta}} (\tau)}{\partial z_1}^\top, \cdots, \frac{\partial \hat{{\beta}} (\tau)}{\partial z_q}^\top \right) \begin{bmatrix}
\nabla_{\theta_1} g_1({w}_1, {\theta_1})^\top(\check{{\theta}}_1 - {\theta}_1) \\
\vdots \\
\nabla_{\theta_q} g_q({w}_q, {\theta_q})^\top(\check{{\theta}}_q - {\theta}_q)
\end{bmatrix} + o_p(1),
\end{align*}
where the last equality follows from an expansion for $g_h(\cdot)$ for $h = 1, \cdots, q$.

Using the same steps as above,  we have the following representation
\begin{equation}
\label{eq:Bahadur_rep2}
\sqrt{n}(\check{{\beta}}(\tau) - \hat{{\beta}}(\tau)) = \sum_{j=1}^q \left[- D_1^{-1}D_{12}^h \sqrt{n}(\check{{\theta}}_j - {\theta}_j) \right] + o_p(1).
\end{equation}
For any $h = 1, \cdots, q$:  $\sqrt{n}(\check{{\theta}}_h - {\theta}_h) \overset{d}{\to} N(0, V_h)$ by recursive implementation  and $D_1^{-1}D_{12}^h$ is bounded by Assumption \ref{assump1}-(A4), we have that $-D_1^{-1} D_{12}^h \sqrt{n} (\check{{\theta}}_h - {\theta}_h ) \overset{d}{\to} N(0, D_1^{-1}D_{12}^h V_h D_{12}^{h^\top} D_1^{-1})$. Let $\check V=Asym.Var\left(\sum_{j=1}^q \left[- D_1^{-1}D_{12}^h \sqrt{n}(\check{{\theta}}_j - {\theta}_j) \right]\right)$, therefore, we have that
\begin{equation}
\label{eq:n_first} 
\sqrt{n}(\check{{\beta}}(\tau) - \hat{{\beta}}(\tau)) \overset{d}{\to} N\left(0, \check V\right).
\end{equation}

Defining now $\check{{\theta}}_h - {\theta}_h = \frac{1}{n} \sum_{i = 1}^n{r}^h_i ({\theta}_h)+o_p(1)$ as the corresponding Bahadur representation for the parameters in the generated regressors in the above equation, we obtain that
\begin{align*}
    Cov & \left(\sqrt{n}(\hat{{\beta}} - {\beta}), \sqrt{n}(\check{{\beta}} - \hat{{\beta}})\right) \\
    &= - D_1^{-1} E \left[ \frac{1}{n}  \sum_{i = 1}^n  \psi(y_i - {z}_i {\beta})  f(0 | {z}_i) \sum_{h = 1}^q \beta_h (\tau) {z}_i^\top \nabla_{\theta_h}  g_h({w}_{hi},{\theta}_h)^\top {r}_i^h ({\theta}_h)  {z}_i \right] D_1^{-1} \\
    &= - D_1^{-1}  M D_1^{-1} +o_{p}(1)
\end{align*}
where $M = \mbox{plim}_{n \to \infty} \frac{1}{n} \sum_{i = 1}^n \left\{ \psi(y_i - {z}_i {\beta})  f(0 | {z}_i) \sum_{h = 1}^q \beta_h {z}_i^\top \nabla_{\theta_h}  g_h({w}_{hi},{\theta}_h)^\top  r^h_i ({\theta}_h)  {z}_i \right\}$.\\
Finally, combining the above covariance term with \eqref{eq:n_second} and \eqref{eq:n_first} with \eqref{eq:n_betahat}, we obtain
\begin{equation*}
\sqrt{n}(\check{{\beta}} - {\beta} )
 \overset{d}{\to} N\left(0, \tau_j(1 - \tau_j)D_1^{-1}D_0D_1^{-1} + \check V - 2 D_1^{-1} M D_1^{-1} \right).
\end{equation*}

\vspace{.5cm}

\noindent\textbf{Step 2. Smoothing $z_j(\beta_{-j}^+)$}

In Step 1 we assume that previous estimation of the QR coefficients enter the generated regressors $\hat z_j(\beta_{-j}^+)$ and produce a differentiable function. This is not the case because $\hat z_j$ contains indicator functions. 

Now consider replacing $\check \beta$ by $\tilde \beta$ and $\hat z$ by $\tilde z$. Here, $\tilde \beta$ is the QR estimator where the generated regressors have the smoothing conditions given in Assumption \ref{assump2}. By the conditions in the same assumption, for all $h=1,...,q$,

\begin{equation}
D_{12}^h \stackrel{p}{\rightarrow} E[ f(0 | {z}_i)\beta_h {z}_i^\top G_h^*({w}_{hi})],\ n\rightarrow\infty,
\end{equation}
where $G_h^*$ is the $h$-th component of $G^*$ defined in Assumption \ref{assump2}.

 \end{proof}
 
\newpage

\bibliographystyle{econometrica}
\bibliography{mqr} 

\end{document}